\documentclass[fleqn,usenatbib]{mnras}

\usepackage{newtxtext,newtxmath}

\usepackage[T1]{fontenc}

\DeclareRobustCommand{\VAN}[3]{#2}
\let\VANthebibliography\thebibliography
\def\thebibliography{\DeclareRobustCommand{\VAN}[3]{##3}\VANthebibliography}

\usepackage{graphicx}	
\usepackage{amsmath}	
\usepackage{booktabs}
\usepackage{algorithm}
\usepackage{algpseudocode}

\usepackage{tikz}

\newcommand\pU{\mathcal{U}}

\usepackage{soul}

\title[Compression for Ly$\alpha$ cosmology]{Optimal data compression for Lyman-$\alpha$ forest cosmology}

\author[Francesca Gerardi et al.]{
Francesca Gerardi,$^{1}$\thanks{E-mail: francesca.gerardi.19@ucl.ac.uk}
Andrei Cuceu,$^{2}$\thanks{NASA Einstein Fellow, E-mail: cuceu.1@osu.edu}
Benjamin Joachimi,$^{1}$
Seshadri Nadathur$^{3}$
and Andreu Font-Ribera$^{4,1}$
\\
$^{1}$ Department of Physics \& Astronomy, University College London, Gower Street, London WC1E 6BT, UK\\
$^{2}$ Center for Cosmology and Astro-Particle Physics, The Ohio State University, Columbus, Ohio 43210, USA\\
$^{3}$ Institute of Cosmology and Gravitation, University of Portsmouth, Burnaby Road, Portsmouth, PO1 3FX, United Kingdom\\
$^{4}$ Institut de Física d’Altes Energies, The Barcelona Institute of Science and Technology, Campus UAB, 08193 Bellaterra (Barcelona), Spain\\
}

\date{Accepted XXX. Received YYY; in original form ZZZ}

\pubyear{2023}

\begin{document}

\defcitealias{Bourboux_Rich_Font-Ribera_Agathe_Farr_Etourneau_Goff_Cuceu_Balland_Bautista_et_al._2020}{dMdB20}

\label{firstpage}
\pagerange{\pageref{firstpage}--\pageref{lastpage}}
\maketitle

\begin{abstract}

The Lyman-$\alpha$ (Ly$\alpha$) three-dimensional correlation functions have been widely used to perform cosmological inference using the baryon acoustic oscillation (BAO) scale.
While the traditional inference approach employs a data vector with several thousand data points, we apply near-maximal score compression down to tens of compressed data elements.
We show that carefully constructed additional data beyond those linked to each inferred model parameter are required to preserve meaningful goodness-of-fit tests that guard against unknown systematics, and to avoid information loss due to non-linear parameter dependencies.
We demonstrate, on suites of realistic mocks and DR16 data from the Extended Baryon Oscillation Spectroscopic Survey, that our compression approach is lossless and unbiased, yielding a posterior that is indistinguishable from that of the traditional analysis.
As an early application, we investigate the impact of a covariance matrix estimated from a limited number of mocks, which is only well-conditioned in compressed space.

\end{abstract}

\begin{keywords}
cosmological parameters -- large-scale structure of universe --  methods: data analysis
\end{keywords}



\section{INTRODUCTION}\label{sect:score_introduction}

In recent decades, the Lyman-$\alpha$ (Ly$\alpha$) forest gained popularity as a probe of the distribution of matter at redshifts $z>2$. The forest consists of a sequence of absorption lines in high-redshift quasar (QSO) spectra, caused by neutral hydrogen placed along the line-of-sight, and hence it is a tracer of the intergalactic medium (IGM). Therefore, it contains cosmological information, and in particular Lyman-$\alpha$ clustering shows the distinct baryon acoustic oscillations (BAO) feature. This feature was first detected in the Ly$\alpha$ auto-correlation function using the Baryon Oscillation Spectroscopic Survey (BOSS) DR9 data \citep{Busca_Delubac_Rich_Bailey_Font-Ribera_Kirkby_Le_Goff_Pieri_Slosar_Aubourg_et_al._2013, Slosar:2013fi, Kirkby:2013fh}, and subsequently extracted from the Ly$\alpha$ cross-correlation with QSOs using DR11 data \citep{BOSS:2013igd}. 

The Ly$\alpha$ forest auto-correlation and its cross-correlation with quasars have been widely used to place constraints on the cosmological model \citep[e.g.][]{Aubourg:2015,Alam:2017,Cuceu:2019,Alam:2021,Cuceu:2023}. These two correlation functions are typically computed on a 2D grid in comoving coordinates along and across the line-of-sight, resulting in high dimensional data vectors, usually $2500$ long for the auto-correlation and $5000$ for the cross-correlation. However, standard BOSS and eBOSS (\citealt{Bourboux_Rich_Font-Ribera_Agathe_Farr_Etourneau_Goff_Cuceu_Balland_Bautista_et_al._2020}; hereafter dMdB20) Ly$\alpha$ forest analyses have so far focused on extracting cosmological information from the BAO peak, which is well localized to a smaller subset of bins. This means that the vector can be reduced to a smaller dimensionality, encoding the information we wish to capture. Hence, in this context, applying a data compression scheme could be useful to optimize the inference. In addition, the accuracy of the parameter estimates is tightly linked to the covariance matrix of the data vector, under the assumption of a Gaussian likelihood. As the true covariance $\mathbf{\Sigma}$ of the correlation function is inaccessible, standard analyses usually estimate it either from large set of mocks or analytically from models of the covariance matrix \citep{Kitaura2016, Wadekar:2020hax}. In Ly$\alpha$ analyses, producing mocks can be a highly computationally-expensive process, therefore only a limited number is available, 100 in the case of \citetalias{Bourboux_Rich_Font-Ribera_Agathe_Farr_Etourneau_Goff_Cuceu_Balland_Bautista_et_al._2020}. However, if the number of samples is significantly lower than the number of data points, the estimate of the covariance is singular and has no inverse \citep{Hartlap:2006kj,PhysRevD.88.063537, Taylor_Joachimi_2014, Sellentin_Heavens_2015, Percival_Friedrich_Sellentin_Heavens_2021}. 

In the eBOSS DR16 analysis, the covariance matrix $\mathbf{C}$ is computed via the sub-sampling method, which, given some dataset, consists of computing the covariance of correlation functions obtained in individual subsamples of the sky. Despite being larger ($\sim 800$) than the number of mocks (100), the number of subsamples is still lower than the number of data points (2500-5000); hence, the covariance matrix must be tested. Alternatively, in the same analysis, the authors computed a Gaussian covariance matrix using the Wick approximation \citep{BOSS:2014hwf} and used it to benchmark the covariance computed from the sub-sampling method. The accuracy of the covariance matrix would increase by alleviating the mismatch between the number of bins and the number of mocks. This can be done by applying a data compression algorithm and evaluating the (compressed) data covariance matrix in a new space characterized by a lower dimensionality. In particular, given the available set of a hundred mocks, we reduce each of them to a set of compressed data vectors and compute a newly defined mock sample covariance, which is a good estimator of the true covariance, given that the length of the compressed data vector is now much smaller than the number of mocks. Then, a comparison between the covariance matrix of the data, mapped into the compressed space, and the mock sample covariance, obtained from the compressed vector, can clarify whether there has been an underestimation or overestimation of the contours in the standard analyses. Moreover, we are interested in obtaining a more sensitive goodness of fit test. The length of Ly$\alpha$ correlation data vectors is of the order of $\mathcal{O}(10^3)$, which could easily hide any bad fit in a subset of the data. By reducing the dimensionality of the data vector through compression, we wish to obtain a test that would highlight when a few important points are off.

Driven by these optimization problems, we perform the inference analysis on realistic Ly$\alpha \times$Ly$\alpha$ auto- and Ly$\alpha \times$QSO cross-correlation functions in a data compression framework. The compression algorithm we use is \textit{score compression} \citep{Alsing_score}, under the hypothesis of a Gaussian likelihood (and hence analogous to the Multiple Optimised Parameter Estimation and Data compression (MOPED) scheme; see \citealt{heavens00}). By construction, the dimensionality of the compressed data vector will be equal to the number of parameters we wish to keep information of, namely $\mathcal{O}(10)$. 

The paper is structured as follows. We start in Sect. \ref{sect:score_method} by outlining the method, explaining the computation of the covariance matrix, and introducing the modelling and the basic idea behind score compression. We proceed in Sect. \ref{sect:score_results} by testing the compression algorithm against loss of information, comparing the inferred posterior distribution for our sampled parameters in the traditional and compressed frameworks. In Sect. \ref{sect:results_testcovariance},  we compare the constraining power of the original estimated covariance matrix against the mock-to-mock covariance. We then perform goodness of fit tests in the compressed framework in Sect. \ref{sect:score_GOODNESS_FIT}. Throughout the analysis a tight prior on the BAO parameters is imposed to overcome the problem of the non-linear relation between these and their corresponding summary statistics components. We relax the prior constraint, and hence made the analysis more generalizable, by extending the framework as described in Sect.~\ref{sect:fiducials_ensemble}. An application of our new framework to eBOSS DR16 data is presented in Sect.~\ref{sect:score_eboss}. Conclusions are drawn in Sect. \ref{sect:score_conclusions}.

Making sure the analysis is both optimized and reliable is key to interpret the vast amount of Ly$\alpha$ forest data which will become available from the Dark Energy Spectroscopic Instrument (DESI).

\section{METHOD}\label{sect:score_method}

Generically referring to the Ly$\alpha$ auto- and cross-correlations as the data vectors, the goal of this work is to study data compression in the context of Ly$\alpha$ forest 3D analyses. In particular, this means compressing the data down to a set of summary statistics $\mathbf{t}$, which will encode into a shorter vector the information we are interested in. As we have just seen, this also benefits the computation of the covariance matrix. The new `compressed' framework is tested against the traditional analysis while performing parameter inference. To evaluate posterior distributions we use the nested sampler \textsc{Polychord} \citep{Handley_Hobson_Lasenby_2015a, Handley_Hobson_Lasenby_2015b}. 

We start in  Sect. \ref{subsect:score_mocks} by introducing the mocks used in this analysis, with a focus on the computation of the covariance matrix. We then describe the modelling of the Ly$\alpha \times$Ly$\alpha$ and the cross Ly$\alpha \times$QSO power spectra in Sect. \ref{subsect:score_method_model}, as implemented in \textsc{vega}\footnote{\url{https://github.com/andreicuceu/vega}} \citep{Cuceu:2022brl}, and the set of randomly generated \textit{Monte Carlo realizations} of the correlation function in Sect.~\ref{subsect:score_method_montecarlo}. In Sect. \ref{subsect:score_method_compression} we finally outline the compression method used, namely \textit{score compression}.

\subsection{Synthetic data vector and covariance} \label{subsect:score_mocks}

In this work we use a set of 100 realistic Ly$\alpha$ mocks, with and without contaminants, which were produced for the Ly$\alpha$ eBOSS DR16 analysis \citep{Bourboux_Rich_Font-Ribera_Agathe_Farr_Etourneau_Goff_Cuceu_Balland_Bautista_et_al._2020}. The synthetic Ly$\alpha$ transmitted fluxes are produced using the \textsc{CoLoRe} \citep{Ram_rez_P_rez_2022} and \textsc{LyaCoLoRe} \citep{Farr_2020} packages, from the same cosmology for all the mocks. Synthetic quasar spectra are then generated given some astrophysical and instrumental prescriptions, and contaminants are added if requested. Then the mocks run through the same analysis pipeline (\textsc{picca}\footnote{\url{https://github.com/igmhub/picca}}) as the real data, resulting in measured auto- and cross-correlation functions (\citetalias{Bourboux_Rich_Font-Ribera_Agathe_Farr_Etourneau_Goff_Cuceu_Balland_Bautista_et_al._2020}). These are derived from computing the correlation function in each HEALPix\footnote{https://healpix.sourceforge.io} \citep{Gorski:2005} pixel --- about 880 pixels (subsamples) for the eBOSS footprint (NSIDE=16) --- and evaluating the mean and covariance over the full set of pixels of the mock, to be then assigned to the entire survey. In this way, for every \textit{i}-th mock, there will be a measurement of both the correlation function and the covariance matrix $\boldsymbol{C}_i$, which will be only an estimate of the true covariance $\boldsymbol{\Sigma}$ as mentioned above. In each subsample, the correlation has a size of either 2500 ($\xi_{\mathrm{auto}}$) or 5000 ($\xi_{\mathrm{cross}}$) bins, hence the number of subsamples (880 pixels) is significantly lower than the number of data points (2500 or 5000). This means that the covariance should be singular, however off-diagonal elements of the correlation matrix are smoothed to make it positive definite (\citetalias{Bourboux_Rich_Font-Ribera_Agathe_Farr_Etourneau_Goff_Cuceu_Balland_Bautista_et_al._2020}).

Finally, given the same hundred mocks, it is possible to define a \textit{stack} of them. In particular, the correlation function for the \textit{stack} of mocks is obtained by collecting all the subsamples (for all the hundred mocks), and computing the mean and covariance of the correlation functions computed in each of them, effectively reducing the noise. We will refer to the contaminated auto- and cross- mock correlations of the \textit{stack} as \textit{stacked correlations}.

In this analysis, we use the same scale cuts as in eBOSS DR16 \citep{Bourboux_Rich_Font-Ribera_Agathe_Farr_Etourneau_Goff_Cuceu_Balland_Bautista_et_al._2020}, assuming $r_{\rm{min}} = 10 \rm{h^{-1}Mpc}$, up to $r_{\rm{max}} = 180 \rm{h^{-1}Mpc}$. The effective redshift of the correlation functions is $z_{\rm{eff}} = 2.3$.

\begin{table*}
\centering
\begin{tabular}{l c c c c c c}
\hline
 &   &  & \multicolumn{2}{|c|}{Testing the framework (\textit{stacked})} & \multicolumn{2}{|c|}{Testing the covariance (single mock)} \\ \cmidrule(lr){4-5}\cmidrule(lr){6-7}
Parameter &  Fiducial & Prior & Traditional & Compression & Original covariance & Mock-to-mock covariance \\
\hline
$\alpha_{\parallel}$ & $1.00$ & $\pU(0.88, 1.14)$ & $ 1.000\pm 0.002 $ & $ 1.000\pm 0.002 $ & $1.003 \pm 0.019$ & $1.003\pm 0.019$\\
$\alpha_{\perp}$ & $1.01$ & $\pU(0.88, 1.14)$ & $ 1.004\pm 0.003 $ & $ 1.004\pm 0.003 $  & $1.002 \pm 0.027$ & $1.004^{+0.029}_{-0.032}$\\
$b_{\mathrm{Ly}\alpha}$ & $-0.14$  & $\pU(-2,0)$ & $ -0.135\pm 0.001 $ & $ -0.135\pm 0.001 $ & $-0.125\pm 0.004$ & $-0.124 \pm 0.006$\\
$\beta_{\mathrm{Ly}\alpha}$ & $1.41$  & $\pU(0,5)$ &  $ 1.47\pm 0.01 $ & $ 1.47\pm 0.01 $ & $1.67^{+0.07}_{-0.08}$ & $1.68^{+0.09}_{-0.10}$\\
$b_{\mathrm{QSO}}$ & $3.81$ &  $\pU(0,10)$ & $ 3.80\pm 0.01 $ & $ 3.80\pm 0.01 $ & $3.82\pm 0.08$ & $3.81\pm 0.07$ \\
$\beta_{\mathrm{QSO}}$ & $0.25$ & $\pU(0,5)$ & $ 0.25\pm 0.01 $ & $ 0.25\pm 0.01 $ & $0.27\pm 0.04$  & $0.27^{+0.03}_{-0.04}$\\
$\sigma_{\mathrm{v}} ({\mathrm Mpc}/h)$ & $2.87$ & $\pU(0,15)$  &  $ 2.82\pm 0.04 $ & $ 2.82\pm 0.04 $ & $3.22^{+0.32}_{-0.28}$ & $3.24\pm 0.26$\\
$\sigma_{\parallel, \mathrm{sm}}$ & $2.05$ & $\pU(0, 10)$ & $ 2.08\pm 0.01 $ & $ 2.08\pm 0.01 $ & $2.10\pm 0.09$ & $2.10^{+0.09}_{-0.08}$\\
$\sigma_{\perp, \mathrm{sm}}$ & $2.35$ & $\pU(0, 10)$ & $ 2.33\pm 0.01 $ & $ 2.33\pm 0.01 $ & $2.23\pm 0.11$ & $2.21\pm 0.11$\\
\hline
$b_{\mathrm{HCD}} [ \times 10^{-2} ]$ & $-1.70$ & $\pU(-20,0)$ & $ -2.12\pm 0.08 $ & $ -2.13\pm 0.07 $  & $-2.98\pm 0.54$  & $-3.06\pm 0.68$ \\
$\beta_{\mathrm{HCD}}$ & $1.57$ & $\mathcal{N}(0.5, 0.09)$ & $ 0.86\pm 0.06 $ & $ 0.86\pm 0.06 $  & $0.50\pm 0.09$ &  $0.50\pm 0.09$\\
$b_{\eta, \mathrm{SiII(1260)}}[ \times 10^{-3} ]$ & $-0.58$ & $\pU(-50, 50)$ & $ -0.59\pm 0.04 $ & $ -0.59\pm 0.04 $  & $-0.83\pm 0.33$ & $-0.88\pm 0.37$\\
$b_{\eta, \mathrm{SiII(1193)}}[ \times 10^{-3} ]$ & $-1.12$ & $\pU(-50, 50)$ & $ -1.09\pm 0.03 $ & $ -1.09\pm 0.03 $  & $-0.83\pm 0.27$ & $-0.84\pm 0.28$\\
$b_{\eta, \mathrm{SiIII(1207)}}[ \times 10^{-3} ]$ & $-1.75$ & $\pU(-50, 50)$ & $ -1.64\pm 0.03 $ & $ -1.63\pm 0.03 $  & $-1.54\pm 0.31$ & $-1.52\pm 0.30$\\
$b_{\eta, \mathrm{SiII(1190)}}[ \times 10^{-3} ]$ & $-1.00$ & $\pU(-50, 50)$ & $ -1.00\pm 0.03 $ & $ -1.00\pm 0.03 $  & $-0.75\pm 0.27$ & $-0.75\pm 0.29$\\
\end{tabular}
\caption{Full set of sampled parameters, alongside with the fiducial values used to compute the summary statistics (see Eq.~\ref{eqn:score_formula}), priors and the 1-D marginals (68\% c.l.). Uniform ($\pU$) priors adopted for the sampling procedure, while we assign a Gaussian prior on $\beta_{\rm{HCD}}$, where by notation the Gaussian distribution $\mathcal{N}(\mu,\sigma)$ has mean $\mu$ and standard deviation $\sigma$. Results are split into `Testing the framework (\textit{stacked})' and `Testing the covariance (single mock)', which respectively refer to the setup in Sect.~\ref{sect:score_results} and Sect.~\ref{sect:results_testcovariance}. The former set of results shows the comparison between the traditional and the compressed inference pipelines using the \textit{stacked} auto- and cross-correlation mocks, while the second between the compressed method using either the original covariance $\boldsymbol{C}$ (which is mapped into the compressed space) or the mock-to-mock covariance $\boldsymbol{C_t}$, for a single mock.} 
\label{tab:score_fulltable}
\end{table*}

\subsection{Modelling and parameter space}\label{subsect:score_method_model}

To model the Ly$\alpha$ correlation functions we follow Eq.~(27) of \citet{Bourboux_Rich_Font-Ribera_Agathe_Farr_Etourneau_Goff_Cuceu_Balland_Bautista_et_al._2020}, while applying the same prescriptions as in \cite{Gerardi2022}. Given a certain cosmological model and a corresponding isotropic linear matter power spectrum $P(k,z)$, the Ly$\alpha$ auto and Ly$\alpha$-QSO cross power spectra are computed as
\begin{align}
    P_{\rm{Ly}\alpha}(k,\mu_k,z) = & b_{\rm{Ly}\alpha}^{2} \left( 1+\beta_{\rm{Ly}\alpha}\mu_k^2 \right)^{2} F_{\rm{nl, Ly}\alpha}^2(k,\mu_k)P(k,z) \label{eqn:score_lyalyapower} \; ; \\
    P_{\times}(k,\mu_k,z) = & b_{\rm{Ly}\alpha} \left( 1+\beta_{\rm{Ly}\alpha}\mu_k^2 \right) \nonumber\\
    &  \times b_{\rm{QSO}} \left(1+\beta_{\rm{QSO}}\mu_k^2 \right) F_{\rm{nl,QSO}}(k_{\parallel})P(k,z) \; , \label{eqn:score_lyaqsopower}
\end{align}
where $\mu_k = k_{\parallel}/k$, with $k$ and $k_{\parallel}$ the wave vector modulus and its line-of-sight component, respectively. On one hand, the Ly$\alpha \times$Ly$\alpha$ power spectrum in Eq.~(\ref{eqn:score_lyalyapower}) depends on the Ly$\alpha$ forest linear bias $b_{\rm{Ly}\alpha}$ and RSD parameter $\beta_{\rm{Ly}\alpha} = \frac{b_{\rm{\eta, Ly}\alpha} f(z)}{b_{\rm{Ly}\alpha}}$, where $b_{\rm{\eta, Ly}\alpha}$ is an extra unknown bias, the velocity divergence bias, and $f(z)$ the logarithmic growth rate. The $F_{ \rm{nl,Ly}\alpha}$ term accounts for non-linear corrections \citep{Arinyo-i-Prats:2015vqa}. On the other hand, the quasar parameters that contribute to the Ly$\alpha \times$QSO power spectrum in Eq.~(\ref{eqn:score_lyaqsopower}) are the quasar linear bias $b_{\rm{QSO}}$ and the redshift-space distortions (RSD) term $\beta_{\rm{QSO}} = f(z)/b_{\rm{QSO}}$. Finally, we model non-linear effects of quasars and redshift errors following \citet{Bourboux_Rich_Font-Ribera_Agathe_Farr_Etourneau_Goff_Cuceu_Balland_Bautista_et_al._2020}, using a Lorentzian function 
\begin{equation}
    F_{\rm{nl, QSO}}(k_{\parallel}) =\left[ 1+ \left( k_{\parallel}\sigma_{\rm{v}} \right)^2 \right]^{-1/2}   \; ,
\end{equation}
where $\sigma_{\rm{v}}$ is the velocity dispersion.

The power spectra in Eqs.~(\ref{eqn:score_lyalyapower}-\ref{eqn:score_lyaqsopower}) only account for Ly$\alpha$ flux and in reality this is also contaminated by absorption lines due to heavy elements, generally referred to as metals, and high column density (HCD) systems \citep{Bautista:2017,Andreu_HCD}. Let us first focus on the modelling of the HCDs. \cite{Andreu_HCD} showed their broadening effect along the line-of-sight can be modeled at the level of new effective Ly$\alpha$ bias and RSD parameters
\begin{align}
    b^{'}_{\mathrm{Ly}\alpha} &= b_{\mathrm{Ly}\alpha} + b_{\mathrm{HCD}}F_{\mathrm{HCD}}(k_{\parallel}) ~,\\
    b^{'}_{\mathrm{Ly}\alpha} \beta^{'}_{\mathrm{Ly}\alpha} &= b_{\mathrm{Ly}\alpha} \beta_{\mathrm{Ly}\alpha} + b_{\mathrm{HCD}} \beta_{\mathrm{HCD}} F_{\mathrm{HCD}}(k_{\parallel}) ~,
\end{align}
with $b_{\mathrm{HCD}}$ and $\beta_{\mathrm{HCD}}$ being the linear bias and RSD parameters. $F_{\mathrm{HCD}}(k_{\parallel})$ is a function of the line-of-sight wavenumber, and it is modeled following \citetalias{Bourboux_Rich_Font-Ribera_Agathe_Farr_Etourneau_Goff_Cuceu_Balland_Bautista_et_al._2020}. On the other hand, metals contribute to the final auto- and cross-correlation functions as per
\begin{align}
    \xi^{'}_{\mathrm{auto}} &= \xi_{\mathrm{Ly}\alpha \times \mathrm{Ly}\alpha} + \sum_m \xi_{\mathrm{Ly}\alpha \times m} + \sum_{m_1, m_2} \xi_{m_1 \times m_2} ~,\\
    \xi^{'}_{\mathrm{cross}} &= \xi_{\mathrm{Ly}\alpha \times \mathrm{QSO}} + \sum_m \xi_{\mathrm{QSO} \times m} ~,
\end{align}
where $m$ generically refer to a metal and the sums are performed over all possible metals considered. The modelling of the cross-correlation of a metal with other metals ($\xi_{m_1 \times m_2}$) and with Ly$\alpha$ ($\xi_{\mathrm{Ly}\alpha \times m}$) and QSO ($\xi_{\mathrm{QSO} \times m}$) follows the modelling of the auto- and cross-correlations of the Ly$\alpha$, and each metal line has a linear bias $b_m$ and RSD parameter $\beta_m=b_{\eta,m}f(z)/b_m$. Following \citetalias{Bourboux_Rich_Font-Ribera_Agathe_Farr_Etourneau_Goff_Cuceu_Balland_Bautista_et_al._2020}, we fix all $\beta_m=0.5$, and sample the metal biases.

Based on this modelling, we use the code \textsc{vega} to compute the two-dimensional correlation functions $\boldsymbol{\xi}$. This same code computes both the BAO feature parameters $\{ \alpha_{\parallel},\alpha_{\perp} \}$, which shift the peak along and across the line-of-sight, and the Gaussian smoothing \citep{Farr_2020}, which accounts for the low resolution of the mocks and is parameterized by $ \{ \sigma_{\parallel}, \sigma_{\perp} \}$ smoothing parameters.

At the inference level, the set of sampled parameters is $\boldsymbol{p_{\rm{s}}} = \{\alpha_{\parallel}, \alpha_{\perp},  b_{\rm{Ly}\alpha}, \beta_{\rm{Ly}\alpha}, b_{\rm{QSO}}, \beta_{\rm{QSO}}, \sigma_{\rm{v}}, \sigma_{\parallel}, \sigma_{\perp} \}$, which is extended to include also $\{ b_{\eta, \mathrm{m}}, b_{\mathrm{HCD}}, \beta_{\mathrm{HCD}} \}$ when also fitting for contaminants. In this notation, $b_{\eta, \mathrm{m}}$ is the velocity divergence bias for the metal $\mathrm{m}$ --- here we consider SiII(1260), SiII(1193), SiIII(1207) and SiII(1190).

For all these parameters we choose uniform priors, which are listed in Tab.~\ref{tab:score_fulltable}. The only exception is given by $\beta_{\rm{HCD}}$, for which, following the previous eBOSS DR16 analysis, we impose an informative Gaussian prior.


\subsection{Monte Carlo realizations}\label{subsect:score_method_montecarlo}

We here introduce a different kind of simulated data, which we will later use, defined as \textit{Monte Carlo realizations}. They are correlation functions obtained by adding noise on top of the model, as defined in Sect.~\ref{subsect:score_method_model}. The noise is given by a covariance matrix from one of the hundred mocks correlation that have been seen so far. What this means is that we can imagine every data point to be generated from a normal distribution $\mathcal{N}(\boldsymbol{\xi}, \boldsymbol{C})$, where $\boldsymbol{\xi}$ is the model correlation function and $\boldsymbol{C}$ is given by the covariance of the first realistic mock. Using Monte Carlo simulations comes with two advantages. First, it is possible to generate as many as needed to build any statistics. Secondly, we have control over the model and there will be clear knowledge of the underlying physics.

\subsection{Score compression} \label{subsect:score_method_compression}
To reduce the dimensionality of our datasets we use score compression \citep{Alsing_score}. Given a known form for the log-likelihood function $\mathcal{L}$, this method corresponds to linear transformations of the data, based on the idea of compressing them down to the score function $\boldsymbol{s} = \nabla \boldsymbol{\mathcal{L}}_*$. The components of the compressed vector are the derivatives of the log-likelihood function, evaluated at some fiducial set of parameters $\boldsymbol{\theta}_*$, with respect to the parameters of interest $\boldsymbol{\theta}$. Under the assumptions that the likelihood function is Gaussian and the covariance $\boldsymbol{C}$ does not depend on parameters, from the data $\boldsymbol{d}$ the compressed data vector is obtained as
\begin{equation}\label{eqn:score_formula}
    \boldsymbol{t} = \nabla \boldsymbol{\mu}_*^T \boldsymbol{C}^{-1} (\boldsymbol{d}-\boldsymbol{\mu}_*) \; ,
\end{equation}
where $\boldsymbol{\mu}_*$ is the fiducial model. 
Under these assumptions the compression is identical to the widely used MOPED scheme \citep{heavens00} apart from a bijective linear transformation.

In our case the model corresponds to the correlation function $\boldsymbol{\xi}$, described earlier in Sect.~\ref{subsect:score_method_model}. The corresponding likelihood distribution in compressed space will be then given by
\begin{equation}\label{eqn:likelihood}
    P(\boldsymbol{t}|\boldsymbol{\theta}) = \dfrac{1}{(2\pi)^{\frac{n}{2}}|\boldsymbol{F}|^{\frac{1}{2}}} \mathrm{exp} \left[ -\dfrac{1}{2}[\boldsymbol{t}-\boldsymbol{\mu}_{\boldsymbol{t}}(\boldsymbol{\theta})]^T \boldsymbol{F}^{-1} [\boldsymbol{t}-\boldsymbol{\mu}_{\boldsymbol{t}}(\boldsymbol{\theta})] \right] \; ,
\end{equation}
where $n$ is the length of $\boldsymbol{t}$, $\boldsymbol{\mu}_{\boldsymbol{t}}(\boldsymbol{\theta})$ is the compressed model $\boldsymbol{\mu}$ evaluated at $\boldsymbol{\theta}$, namely $\boldsymbol{\mu}_{\boldsymbol{t}}(\boldsymbol{\theta}) = \nabla \boldsymbol{\mu}_*^T \boldsymbol{C}^{-1} [\boldsymbol{\mu}(\boldsymbol{\theta})-\boldsymbol{\mu}_*]$, and 
\begin{equation}\label{eqn:fisher}
    \boldsymbol{F} = [\nabla \boldsymbol{\mu}_*]^T \boldsymbol{C}^{-1}[\nabla^{T} \boldsymbol{\mu}_*]
\end{equation}
is the Fisher matrix.

When considering both the auto- and cross-correlations, some parameters will be in common; for this reason, there is the need to build a joint summary statistic. If we define independently the Ly$\alpha$ auto- and cross- data vectors, characterized by the covariances $\mathbf{C}_{\rm{auto}}$ and $\mathbf{C}_{\rm{cross}}$ respectively, and given they do not correlate with each other, in the joint analysis the full covariance matrix will be given by
\begin{equation}
    \mathbf{C} =
    \begin{pmatrix}
    \mathbf{C}_{\rm{auto}} & 0 \\
    0 & \mathbf{C}_{\rm{cross}}
    \end{pmatrix} ~~.
\end{equation}
Then the resulting summary statistics vector and Fisher matrix will be respectively obtained as $\boldsymbol{t} = \boldsymbol{t}_{\rm{auto}} + \boldsymbol{t}_{\rm{cross}}$ and $\boldsymbol{F} = \boldsymbol{F}_{\rm{auto}} + \boldsymbol{F}_{\rm{cross}}$.

This compression method is dependent on the choice of the fiducial set of parameters $\boldsymbol{\theta}_*$, which might not be known \textit{a priori}. However, \cite{Alsing_score} suggest iterating over the \textit{Fisher scoring method} for maximum-likelihood estimation
\begin{equation}\label{eqn:iteration_formula}
    \boldsymbol{\theta}_{k+1} = \boldsymbol{\theta}_k + \boldsymbol{F}^{-1}_k \nabla \boldsymbol{\mathcal{L}}_k \; ,
\end{equation}
until convergence of the full set of parameters. How this is done in our particular case is described at the beginning of Sect.~\ref{sect:score_results}. An important note is that this iterative procedure does not take into account parameters priors, which means it can potentially lead to unusual values for those parameters which are not well constrained by the data.

\begin{figure}
    \centering
    \includegraphics[width=0.48\textwidth]{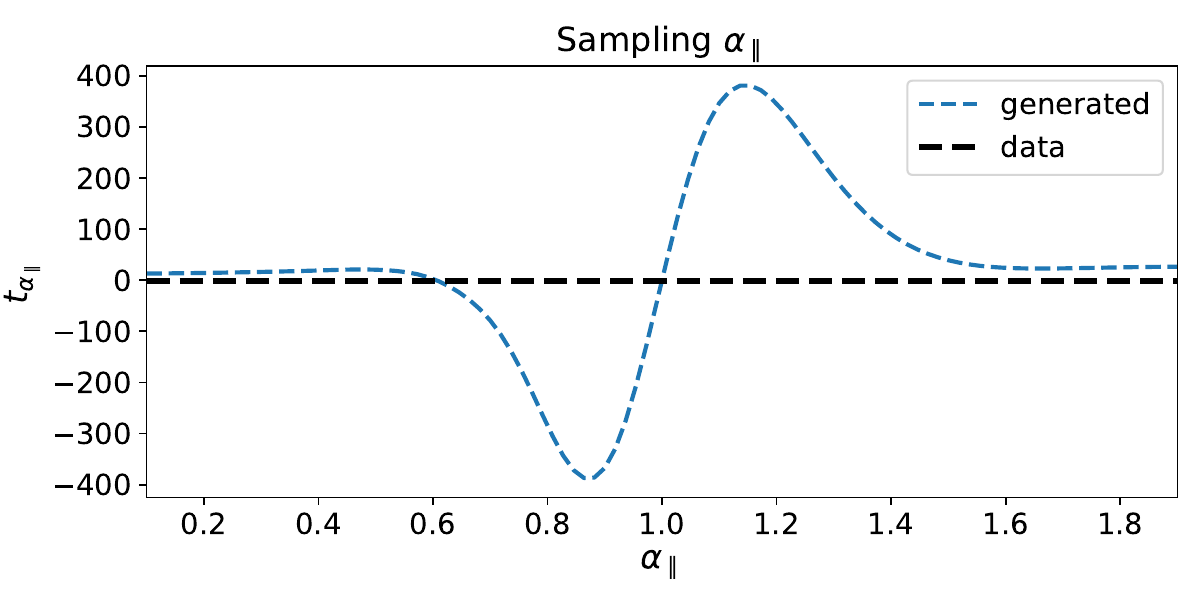}
    \caption{This plot shows the behaviour of the summary component $t_{\alpha_{\parallel}}$ as a function of $\alpha_{\parallel}$, which is the parameter it is related to as per Eq.~(\ref{eqn:score_formula}), against the value of $t_{\alpha_{\parallel}}$ evaluated using $\alpha_{\parallel} = 1.00$ (see Tab. \ref{tab:score_fulltable}), denoted as `data'. The remainder of the parameters are set to the fiducial values listed in Tab~\ref{tab:score_fulltable}. This figure highlights a non-monotonic relationship between the two parameters, which would lead to multiple peaks in the posterior if a tight prior is not imposed.}
    \label{fig:score_apcompressed}
\end{figure}

In computing the score compression components over the parameters $\{ \alpha_{\parallel},\alpha_{\perp} \}$, we realized their relation with their corresponding summary statistics components, namely $\{ \boldsymbol{t}_{\alpha_{\parallel}},\boldsymbol{t}_{\alpha_{\perp}} \}$, was not monotonic, as per Fig. \ref{fig:score_apcompressed}. This is problematic as this means the posterior can have more than one peak \citep{Graff_Hobson_Lasenby_2011} if we sample over the full $[0.01, 1.99]$ interval. We overcame this complexity by imposing a tighter prior on $\{ \alpha_{\parallel},\alpha_{\perp} \}$ at the sampling step. This prior is designed to allow for $\alpha_{\parallel}$ values in between the minimum and maximum of $\boldsymbol{t}_{\alpha_{\parallel}}(\alpha_{\parallel})$. The same prior is imposed on $\alpha_{\perp}$. This tightening does not affect the inference when performed on the correlation function of the \textit{stacked} mock, in which case posteriors are well within this prior, but it reveals to be quite important when evaluating the posteriors on the individual mocks. For this reason, we make sure we provide example results for those mocks whose contours are within the prior range.

Later, in Sect.~\ref{sect:fiducials_ensemble} we will see how we can remove the tight prior constraint by evaluating the summary statistics components associated to $\{ \alpha_{\parallel},\alpha_{\perp} \}$ at multiple fiducial values of the BAO parameters, effectively enlarging the compressed vector.

\begin{figure*}
    \centering
    \begin{tikzpicture}
      \node (img1) {\includegraphics[width=0.78\textwidth, clip=true, trim = 0.47cm 0.5cm 0.35cm 0.3cm]{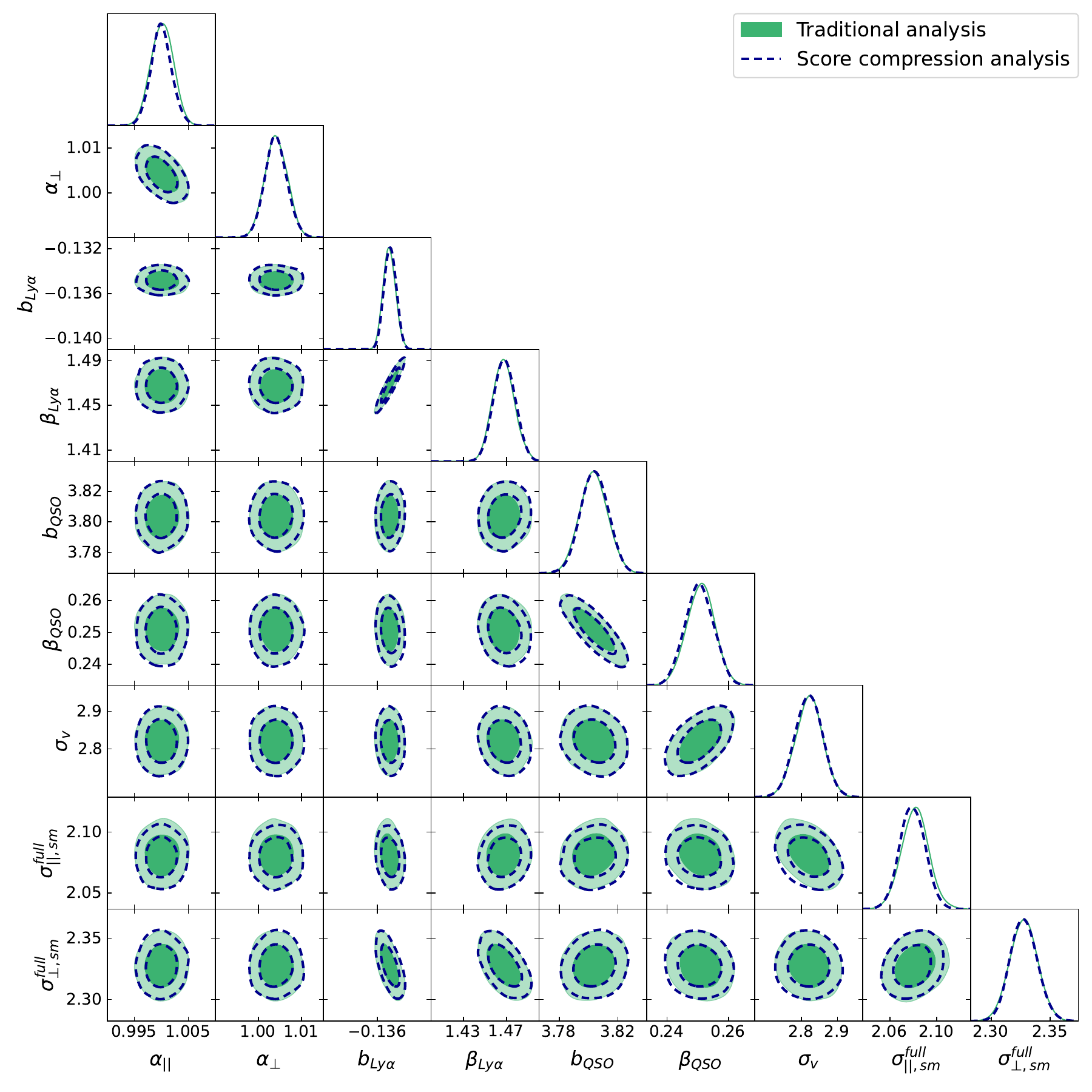}};
      \node (img2) at (6.2, 4.6) {\includegraphics[width=0.55\textwidth, clip=true, trim = 0.49cm 0.5cm 0.35cm 0.3cm]{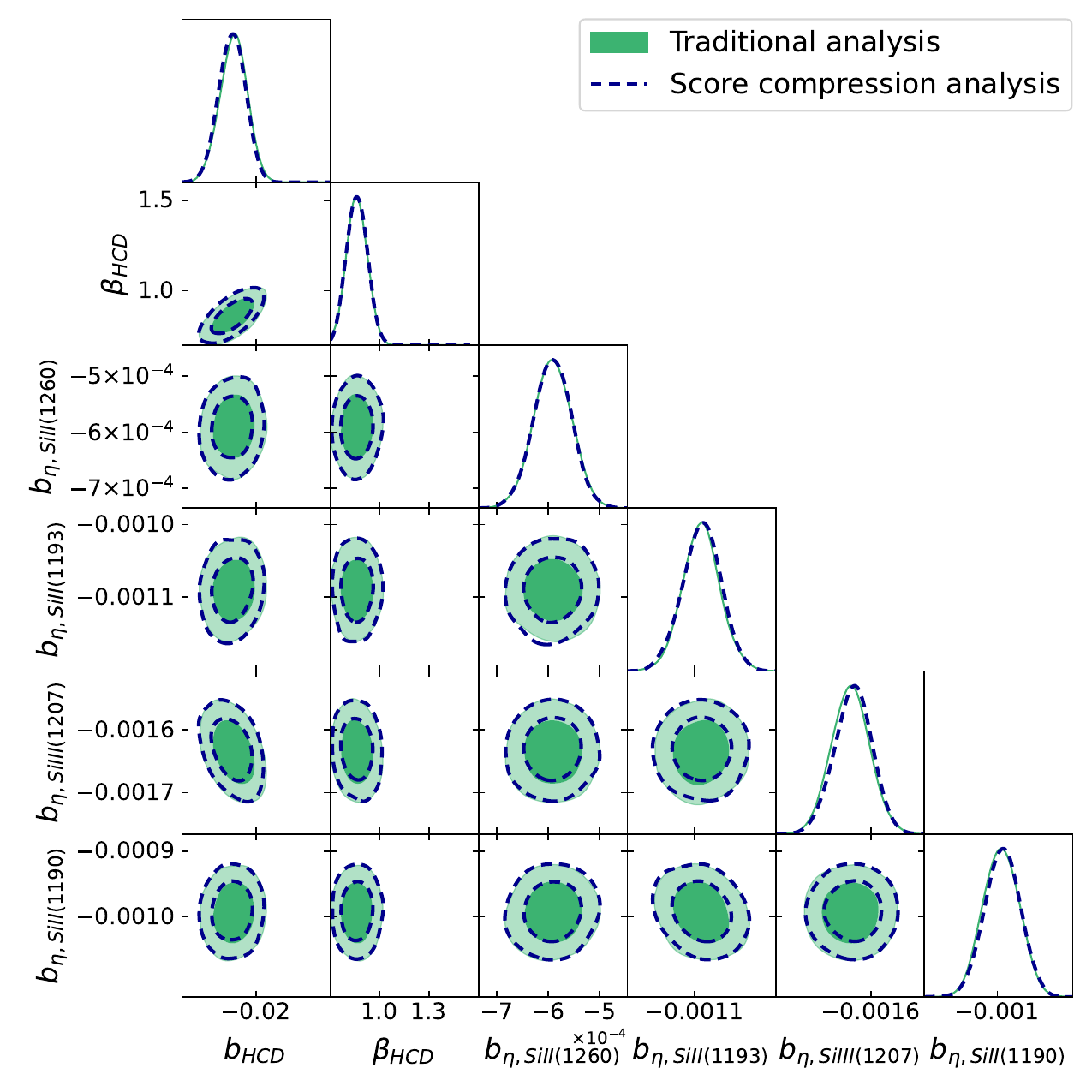}};
    \end{tikzpicture}
    \caption{Triangle plots of the parameters of interest for the  \textit{stack} of correlation functions computed from a set of 100 mocks. Results are split, for presentation purposes only, into the set of standard parameters $\{\alpha_{\parallel}, \alpha_{\perp},  b_{\rm{Ly}\alpha}, \beta_{\rm{Ly}\alpha}, b_{\rm{QSO}}, \beta_{\rm{QSO}}, \sigma_{\rm{v}}, \sigma_{\parallel}, \sigma_{\perp} \}$ (lower left panel) and contaminants parameters $\{ b_{\eta, \mathrm{SiII(1260)}}, b_{\eta, \mathrm{SiII(1193)}}, b_{\eta, \mathrm{SiIII(1207)}}, b_{\eta, \mathrm{SiII(1190)}}, b_{\mathrm{HCD}}, \beta_{\mathrm{HCD}} \}$ (upper right panel). The green contours refer to the results obtained performing the inference using the full uncompressed data vector, which we denote as ‘Traditional analysis’, while the blue dashed refer to the compressed analysis results, denoted as `Score compression analysis'.}
    \label{fig:score_global_posterior}
\end{figure*}

\section{Compression performance}\label{sect:score_results}

In this Section we apply the score compression algorithm, outlined in Sect~\ref{subsect:score_method_compression}, to Ly$\alpha$ auto- and cross-correlations measured from contaminated mocks. The pipeline starts by choosing a fiducial set of parameters for computing the score compressed vector, as per Eq.~(\ref{eqn:score_formula}). The fiducial is obtained by iterating over Eq.~(\ref{eqn:iteration_formula}), with $\boldsymbol{\theta}_0$ given by the best fit of the \textit{stacked} correlation functions. Given this initial guess, we then iterated assigning to $\boldsymbol{\theta}_{k+1}$ the median of the $\boldsymbol{\theta}$ values over the hundred mocks at the $k$-th step. 

The likelihood is assumed to be Gaussian, which has a major impact on the final form of the compressed vector, given that the latter is computed as the gradient of the log-likelihood. Based on previous analyses, we assume the data are normally distributed and this assumption of Gaussianity will also be inherited in the compressed space. In general, when mapping in a compressed space, this property might not be preserved, but given that score compression is a linear transformation, that is the case. We make a consistency check by running the Henze-Zirkler test \citep{henze1990class} for multivariate normality in the compressed space. Intuitively, this test measures the distance between the measured and target (multivariate) distribution, 
and it was shown to perform well in high-dimensional problems. We found that the summary statistics, computed for the hundred mocks at the end of the iterative process, follows a multivariate normal distribution.

Provided the fiducial model and the Gaussianity checks, we first test the compression method on the \textit{stack} of the mocks, with results presented in this Section, and later, in Sect.~\ref{sect:results_testcovariance}, we compute the covariance matrix for the summary statistics over the set of hundred mocks and compare it to the Fisher matrix as defined in Eq.~(\ref{eqn:fisher}). It is important to keep in mind that, when referring to the Fisher matrix, we are simply referring to the mapping of the data covariance matrix $\boldsymbol{C}$ into the compressed space.


To test the score compression algorithm against the traditional approach, for simplicity, we employ both the contaminated auto- and cross- \textit{stacked correlations}, which are almost noise-free. This choice is motivated by the fact that we imposed a tight prior on the $\{ \alpha_{\parallel},\alpha_{\perp} \}$ parameters to overcome the challenges coming from the non-monotonic relationship between these parameters and their corresponding summary statistics components (see Fig.~\ref{fig:score_apcompressed}). Thus, experimenting over less noisy mock data facilitates running the test in a case where it is granted that posteriors will not hit the priors. 

For both the traditional (uncompressed data) and the compressed frameworks we run the \textsc{Polychord} sampler for the auto- and cross- \textit{stacked correlations}, while sampling the full set of 15 model parameters $\{\alpha_{\parallel}, \alpha_{\perp},  b_{\rm{Ly}\alpha}, \beta_{\rm{Ly}\alpha}, b_{\rm{QSO}}, \beta_{\rm{QSO}}, \sigma_{\rm{v}}, \sigma_{\parallel}, \sigma_{\perp} , b_{\eta, \mathrm{SiII(1260)}}, $ $ b_{\eta, \mathrm{SiII(1193)}}, b_{\eta, \mathrm{SiIII(1207)}}, b_{\eta, \mathrm{SiII(1190)}}, b_{\mathrm{HCD}}, \beta_{\mathrm{HCD}} \}$ and results are presented in Fig.~\ref{fig:score_global_posterior}. The two methods agree well with each other, leading to almost identical results. The numerical values of the peaks and $1\sigma$ confidence intervals of the 1d marginals are presented in Tab.~\ref{tab:score_fulltable} as part of the `Testing the framework (\textit{stacked})' set of columns. From the table, it can be noticed that in some cases the fiducial parameters used to compute the compression are not within the $3\sigma$ confidence interval. Despite the fiducial being a first guess, and not necessarily accurate, the contours of the two methods agree well with each other.

We just demonstrated that the score compression inference pipeline leads to the same results as the standard approach. This shows the linearity of the parameters in the model to a good approximation. However, it is important to bear in mind that, in this case, this only holds locally around the fiducial, because of the non-linearity of the components that relate to $\alpha_{\parallel}$ and $\alpha_{\perp}$, on which we imposed a tight prior. 

\section{Testing the covariance matrix}\label{sect:results_testcovariance}

An interesting application of the compression algorithm consists of evaluating the accuracy of the covariance matrix $\boldsymbol{C}$ by comparing it to the mock-to-mock covariance $\boldsymbol{C_{t}}$, which is the covariance matrix of the summary statistics vectors of the set of hundred mocks. We now showcase this application using a single mock. 

We recall that the computation of the standard data covariance happens in a setup where the length of the data vector is larger than the number of samples, which is sub-optimal. The covariance should be singular; however, the off-diagonal elements of the correlation matrix are smoothed to make it positive definite \citep{Bourboux_Rich_Font-Ribera_Agathe_Farr_Etourneau_Goff_Cuceu_Balland_Bautista_et_al._2020}. Reducing the dimensionality of the data vector via score compression allows us to compute a new covariance matrix $\boldsymbol{C_t}$, which has a dimensionality significantly lower than the number of samples used to compute it, given that the new data vector will be $\sim \mathcal{O}(10)$ long. The fact that now the number of mock samples is larger than the number of compressed data points, means that we are now in a framework where the estimated $\boldsymbol{C_t}$ is in principle a better estimator of the true covariance $\boldsymbol{\Sigma}$ in compressed space than $\boldsymbol{F}$, which is obtained by mapping the covariance $\boldsymbol{C}$ into this space. 

We now repeat the same experiment as in Sect.~\ref{sect:score_results} over a single mock and evaluate the posterior using \textsc{Polychord} for the full set of parameters $\{\alpha_{\parallel}, \alpha_{\perp},  b_{\rm{Ly}\alpha}, \beta_{\rm{Ly}\alpha}, b_{\rm{QSO}}, \beta_{\rm{QSO}}, \sigma_{\rm{v}}, \sigma_{\parallel}, \sigma_{\perp} , b_{\eta, \mathrm{SiII(1260)}},$ $b_{\eta, \mathrm{SiII(1193)}}, b_{\eta, \mathrm{SiIII(1207)}}, b_{\eta, \mathrm{SiII(1190)}}, b_{\mathrm{HCD}}, \beta_{\mathrm{HCD}} \}$. This is either done using the original covariance $\boldsymbol{C}$ matrix (mapped into the compressed space, to the Fisher matrix) in the Gaussian likelihood in Eq.~(\ref{eqn:likelihood}) or instead using the mock-to-mock covariance $\boldsymbol{C_{t}}$ adopting a t-distribution as a likelihood function, as proposed in \cite{Sellentin_Heavens_2015}. The latter is of the form of
\begin{equation}\label{eqn:likelihood_ct}
    P(\boldsymbol{t}|\boldsymbol{\theta}) = \dfrac{\Bar{c}_P |\boldsymbol{C_{t}}|^{-1/2}}{1+\frac{[\boldsymbol{t}-\boldsymbol{\mu}_{\boldsymbol{t}}(\boldsymbol{\theta})]^T \boldsymbol{C_{t}}^{-1} [\boldsymbol{t}-\boldsymbol{\mu}_{\boldsymbol{t}}(\boldsymbol{\theta})]}{n_s-1}} \; 
\end{equation}
with
\begin{equation}\label{eqn:cp}
    \Bar{c}_P = \dfrac{\Gamma \left(\dfrac{n_s}{2} \right)}{[\pi(n_s-1)]^{n_t/2}\Gamma \left(\dfrac{n_s-n_t}{2} \right)} \; ;
\end{equation}
where $n_s$ is the number of mocks, $n_t$ is the length of the compressed data vector and $\Gamma$ is the Gamma function.
Once again the choice of the tight prior on both $\{\alpha_{\parallel}, \alpha_{\perp}\}$ affected the choice of the set of mocks in order to run this second experiment. However, the goal of this second experiment is to provide an example case of testing the accuracy of the subsampling estimation of the covariance matrix. If the method is demonstrated to effectively work over some subset of mocks, it is expected that will also be the case for the others.

\begin{figure}
    \centering
    \begin{tikzpicture}
      \node (img1) {\includegraphics[width=0.3\textwidth, clip=true, trim = 0.5cm 0.5cm 0.35cm 0.3cm]{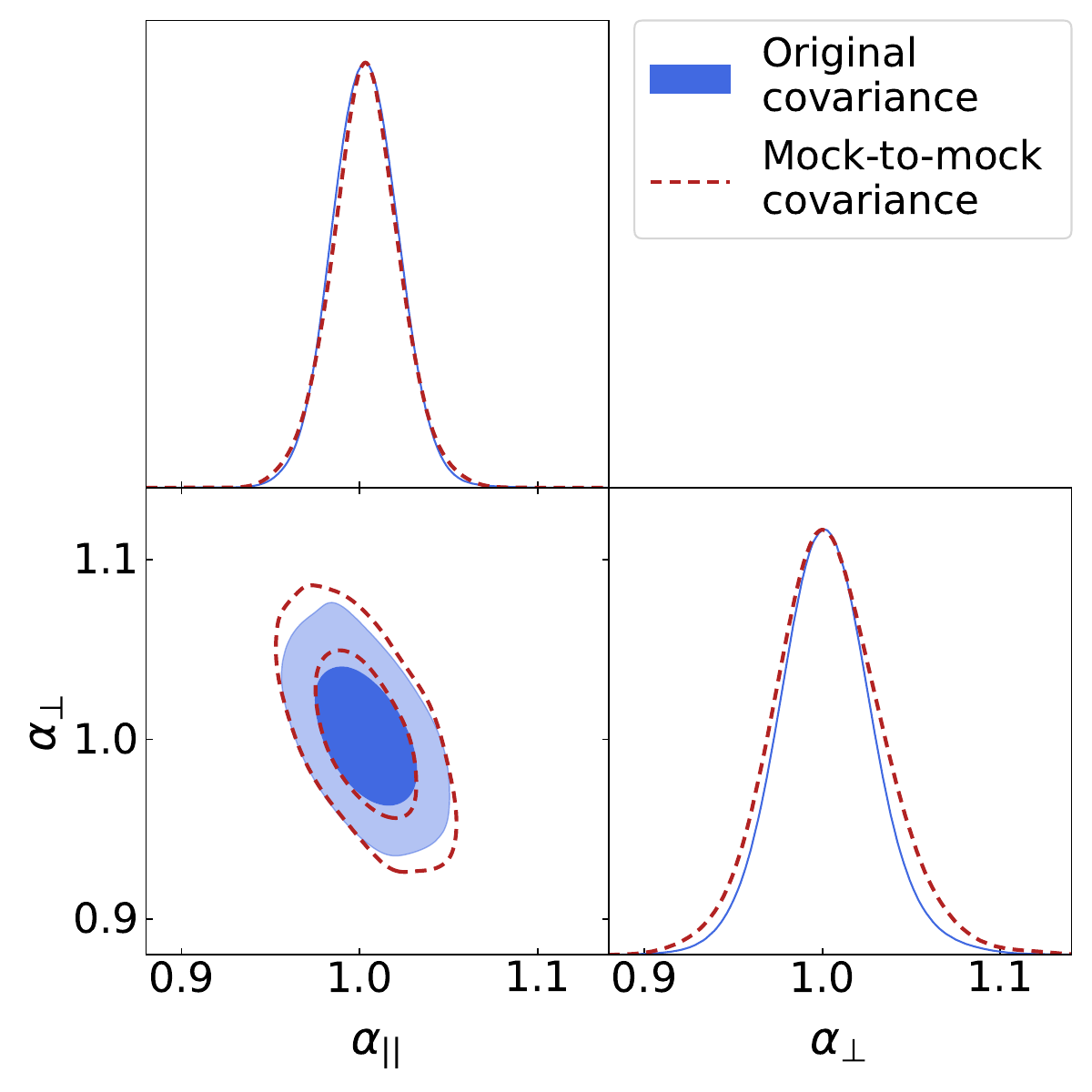}};
      \node (img2) at (3.1, 3.1) {\includegraphics[width=0.3\textwidth, clip=true, trim = 0.4cm 0.5cm 0.35cm 0.3cm]{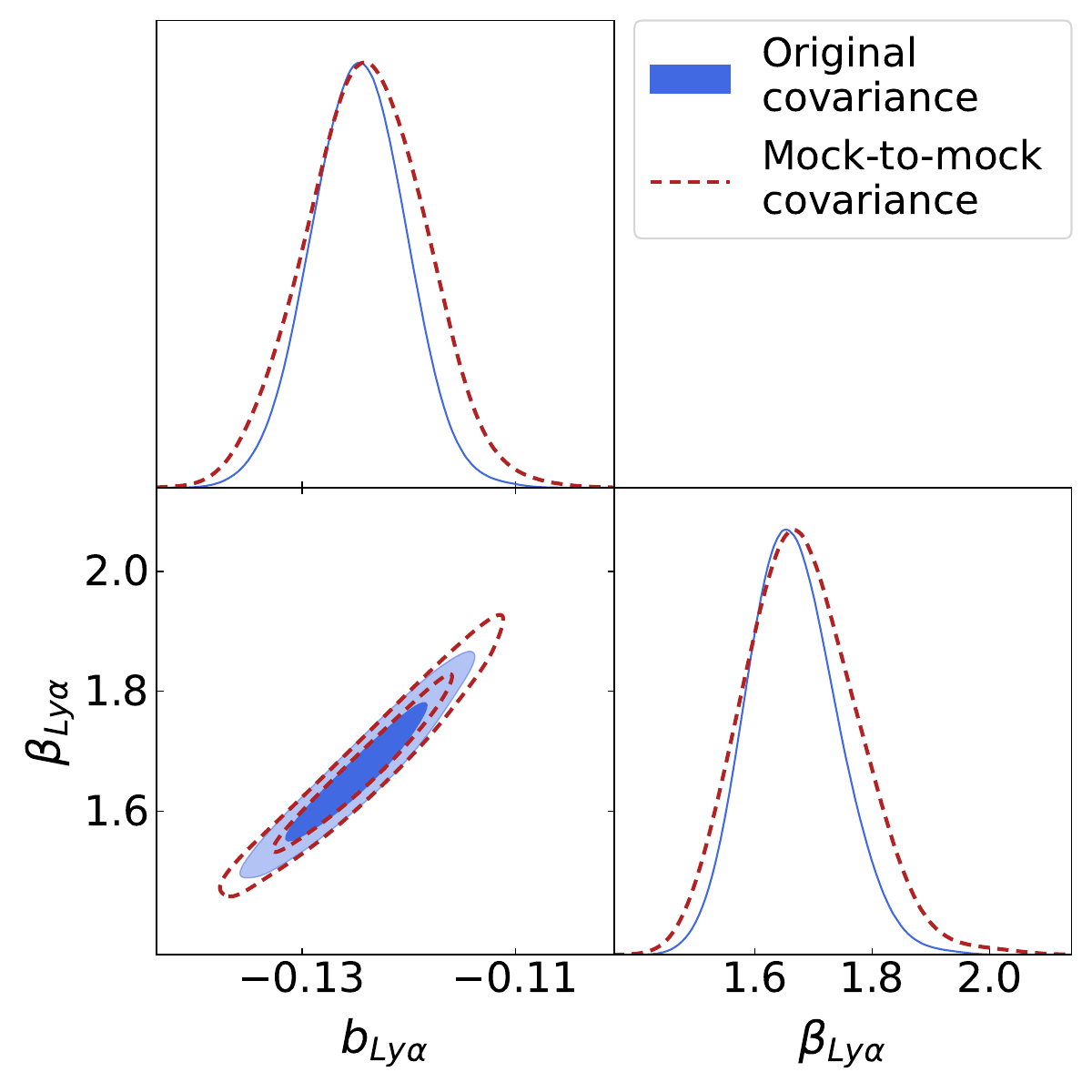}};
    \end{tikzpicture}
    \caption{Triangle plots of the BAO parameters of interest $\{\alpha_{\parallel}, \alpha_{\perp}\}$ and the Ly$\alpha$ parameters $\{b_{\mathrm{Ly}\alpha}, \beta_{\mathrm{Ly}\alpha}\}$ for one set of the Ly$\alpha$ auto- and cross- mock correlations. The blue filled contours refer to the results obtained performing the inference using the original covariance matrix $\boldsymbol{C}$ (mapped into the compressed space) in the likelihood function, and hence are denoted as ‘Original covariance’. On the other hand, the red dashed results, denoted as 'Mock-to-mock covariance', refer to the case in which the mock-to-mock covariance matrix is used instead, while adopting a t-distribution likelihood.}
    \label{fig:score_cov_posterior_apat}
\end{figure}

The results for the BAO parameters $\{\alpha_{\parallel}, \alpha_{\perp}\}$ and the Ly$\alpha$ parameters $\{b_{\mathrm{Ly}\alpha}, \beta_{\mathrm{Ly}\alpha}\}$ are shown in Fig.~\ref{fig:score_cov_posterior_apat}, while the full set is presented in Sect.~\ref{appendix} and listed in Tab.~\ref{tab:score_fulltable} (`Testing the covariance (single mock)' columns). 
In this test case, using the mock-to-mock covariance results in a small enlargement of the posterior for the $\alpha_{\perp}$ parameter: while using the original covariance matrix provides $\alpha_{\perp} = 1.002 \pm 0.027$, the mock-to-mock covariance results in $\alpha_{\perp} = 1.004^{0.029}_{-0.032}$. On the other hand, the Ly$\alpha$ linear bias and RSD parameter absolute errors increase by $50 \%$ and $\sim 25\%$ respectively, with final relative error of about $5-6\%$. The uncertainty of the vast majority of the other parameters agree remarkably well. 

We end this discussion on covariance matrix estimation by noting that the test presented here is meant as a showcase of the usefulness of compressing Ly$\alpha$ forest correlation functions. However, proper testing of the Ly$\alpha$ forest covariance matrices would require a more comprehensive analysis using a larger sample of mocks\footnote{Note also that this kind of analysis heavily relies on mocks being consistent with each other (both in terms of mock production, and in terms of analysis), in order to avoid introducing extra variance.}, and comparison with other estimation methods \citep[see e.g.,][]{Bourboux_Rich_Font-Ribera_Agathe_Farr_Etourneau_Goff_Cuceu_Balland_Bautista_et_al._2020}.

\begin{figure}
    \centering
    \includegraphics[width=0.5\textwidth]{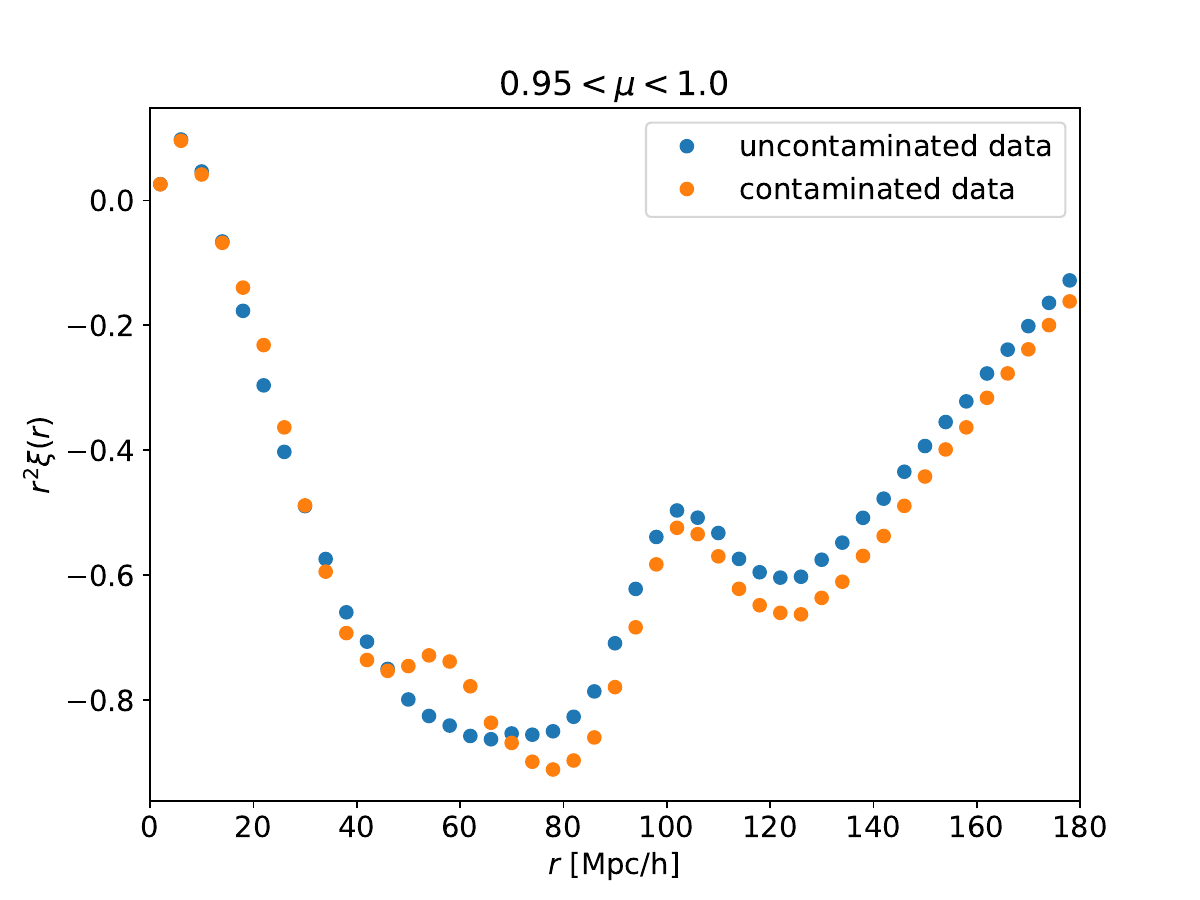}
    \caption{This wedge plot, for $|\mu| = |r_{\parallel}/r|$ between 0.95 and 1.0, shows the effect of adding metals (in orange) to the correlation model $\boldsymbol{\xi}$ without metals (in blue) along the line-of-sight. For simplicity in the $\chi^2$ analysis we do not include contamination coming from HCD, so these features are only the effects of metal lines. Also, in this example, in order to better visualize the difference between the two, we have been generating noise from the covariance matrix of the \textit{stacked} auto-correlation mock.}
    \label{fig:correlations}
\end{figure}

\section{Goodness of fit test}\label{sect:score_GOODNESS_FIT}

\begin{figure*}
    \centering
    \includegraphics[width=0.325\textwidth, clip=true, trim = 0.4cm 0.5cm 0.35cm 0.3cm]{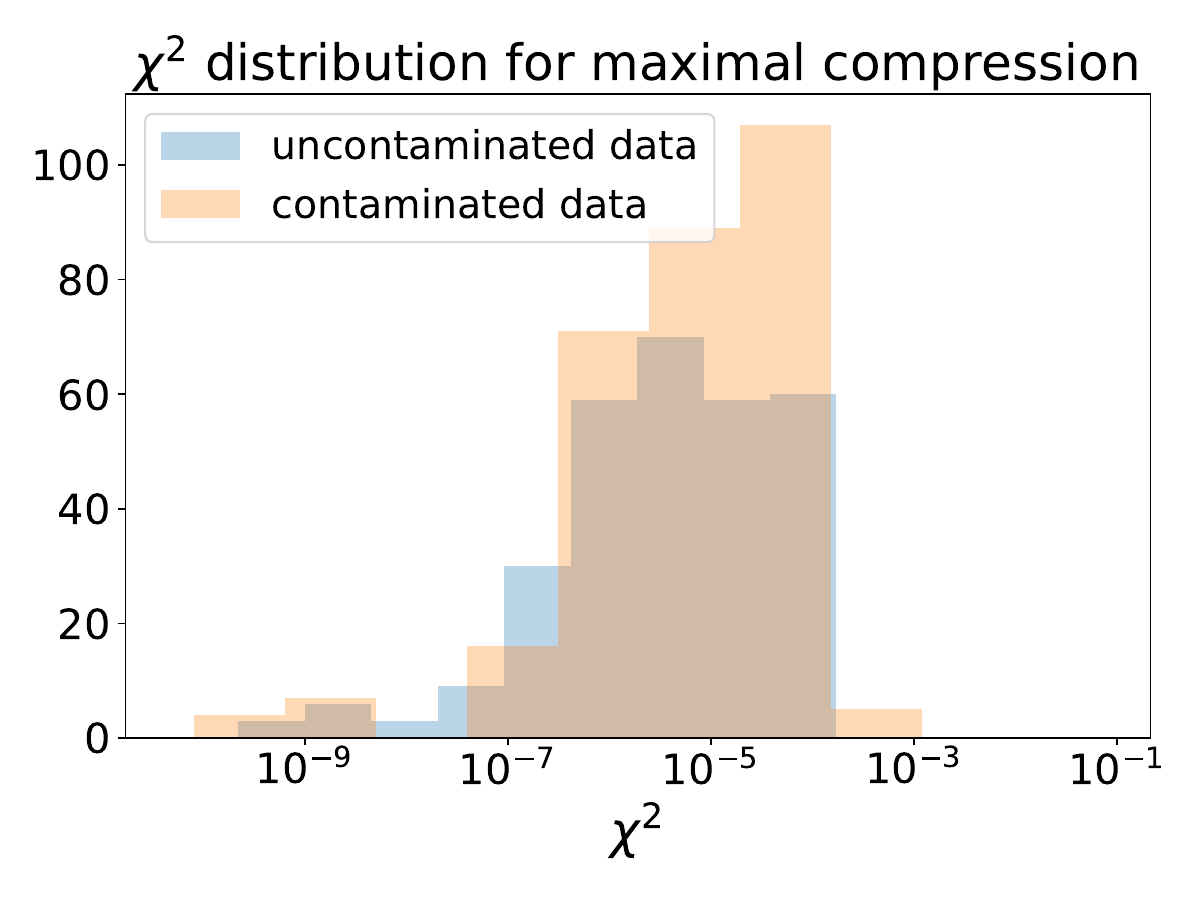}
    \hspace{1cm}
    \includegraphics[width = 0.6\textwidth, clip=true, trim = 0.4cm 0.05cm 0.35cm 0.cm]{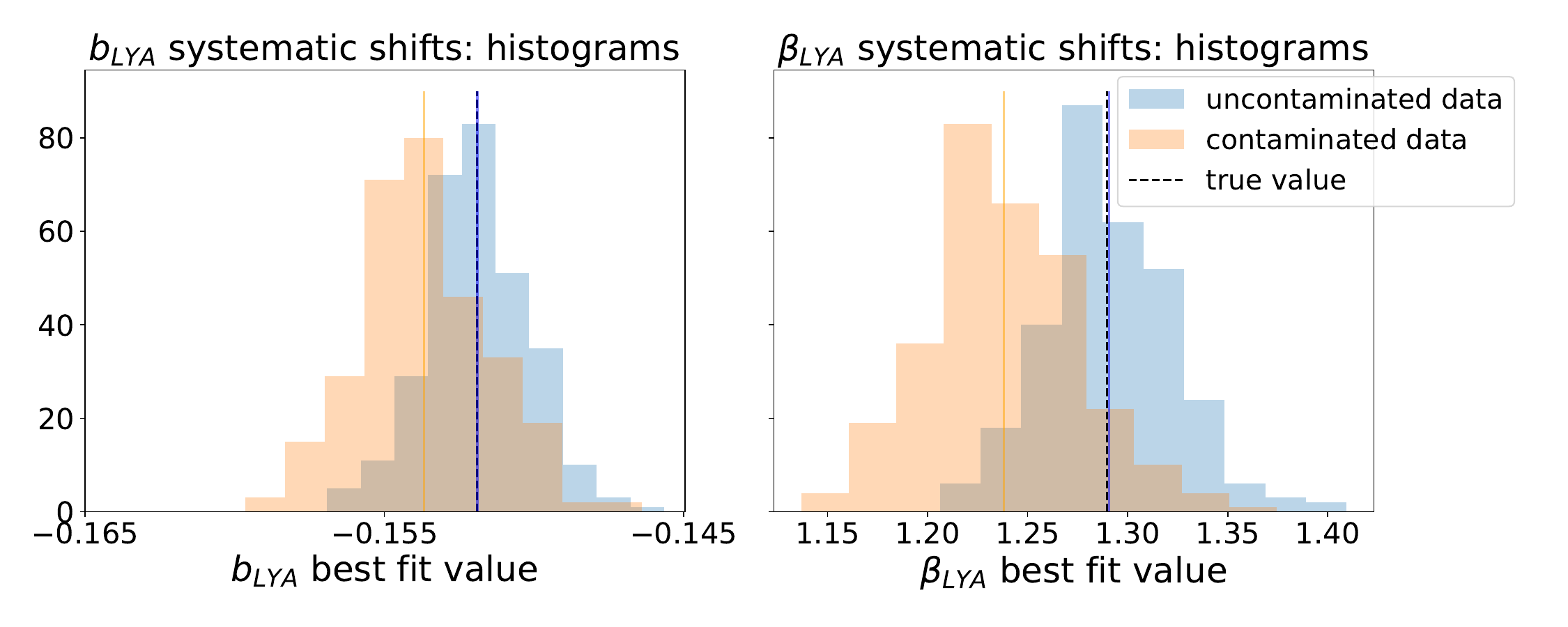}
    \caption{$\chi^2$ histograms (left panel) for the \textit{maximal} compression and corresponding best fit values histograms for the Ly$\alpha$ parameters (right panels), where blue refers to the uncontaminated case and orange to contaminated. In the \textit{maximal} compression setup $\boldsymbol{t} = \boldsymbol{t}_{\mathrm{max}} = \{t_{\alpha_{\parallel}}, t_{\alpha_{\perp}},  t_{b_{\mathrm{Ly}\alpha}}, t_{\beta_{\mathrm{Ly}\alpha}}, t_{\sigma_{\parallel}}, t_{\sigma_{\perp}} \}$. The black dashed lines in the two panels on the right correspond to the true values used to generate the Monte Carlo realisations.} 
    \label{fig:all_cases_chi2}
\end{figure*}

In this section, we make a step forward with respect to the original aim of the work, by considering goodness of fit tests. For Ly$\alpha$ correlation functions, the length of the data vector can go from 2500, considering only the auto-, to 7500 if considering also the cross-correlation. In a context where only $\sim \mathcal{O}(10)$ parameters are sampled, any bad fit for noisy data can be hard to detect. Reducing the dimensionality of the data via score compression, we investigate whether it would be easier for any bad fit to be spotted. Hence, given the results presented in Sect.~\ref{sect:score_results}, we test the robustness of the method against unmodelled effects in the correlation functions, via the $\chi^2$ statistics. 

To this end we test the goodness of fit on contaminated data when metals are not modelled. For simplicity, here we restrict to the Ly$\alpha$ auto-correlation alone and without considering contamination from HCD. The sampled parameters will only be $ \{\alpha_{\parallel}, \alpha_{\perp},  b_{\rm{Ly}\alpha}, \beta_{\rm{Ly}\alpha}, \sigma_{\parallel}, \sigma_{\perp} \}$. Tests are run by constructing the $\chi^2$ distributions over a set of 300 Monte Carlo realizations of the auto-correlation, introduced in Sect.~\ref{subsect:score_method_montecarlo}: for each realization we run a minimizer and evaluate the $\chi^2$ at the best fit.

We considered two main Monte Carlo populations: with and without metal contamination. The difference between the two is shown in the wedge plot of Fig.~\ref{fig:correlations}, which is built by averaging over the values of the correlation function in the `wedge' of the space $\{r_{\parallel},r_{\perp}\}$ identified by values of $|\mu| = |r_{\parallel}/r|$ between 0.95 and 1.0. To generate them we used the best fit values of $ \{\alpha_{\parallel}, \alpha_{\perp},  b_{\rm{Ly}\alpha}, \beta_{\rm{Ly}\alpha}, \sigma_{\parallel}, \sigma_{\perp},  b_{\eta, \mathrm{SiII(1260)}}, b_{\eta, \mathrm{SiII(1193)}},$ $ b_{\eta, \mathrm{SiIII(1207)}}, b_{\eta, \mathrm{SiII(1190)}}\}$ for the contaminated \textit{stacked} Ly$\alpha$ mock auto-correlation, where depending on the population (contaminated or uncontaminated) the metals' parameters were either included or not.

\subsection{Maximal compression}

For both the contaminated and uncontaminated mock data, we apply a compression down to the same number of sampled parameters without including contamination in the modelling, with the summary statistics thus given by $\boldsymbol{t}_{\mathrm{max}} = \{t_{\alpha_{\parallel}}, t_{\alpha_{\perp}},  t_{b_{\rm{Ly}\alpha}}, t_{\beta_{\rm{Ly}\alpha}}, t_{\sigma_{\parallel}}, t_{\sigma_{\perp}} \}$. This is defined as \textit{maximal compression}. In what follows we are interested in learning about the $\chi^2$ distribution for the two Monte Carlo populations. 

We found that for both contaminated and uncontaminated data, the $\chi^2$ distributions are similar, with values of the order of $\mathcal{O}(10^{-10}-10^{-3})$ (left panel of Fig.~\ref{fig:all_cases_chi2}). However, comparing the fits to the contaminated and uncontaminated data, the best-fit parameter values are systematically shifted for some parameters. The distributions of the best-fit values for $b_{\mathrm{Ly}\alpha}$ and $\beta_{\mathrm{Ly}\alpha}$ are shown in the right panels of Fig.~\ref{fig:all_cases_chi2}: for the fits to contaminated data, 80\% and 90\% of the best-fit values respectively for each parameter are below the true value. 

The $\chi^2$ values remain very small for the fits to contaminated data, which indicates that in the compressed space, the model without contaminants still has enough flexibility to perfectly fit the data: the system has zero degrees of freedom, given that we are sampling six parameters, and the compressed data vector has six components. Instead of the mismatch between the model without contaminants and the contaminated data being visible in the form of large $\chi^2$ values, it is manifested through a systematic shift in the recovered parameter values from the truth, which in a realistic data fitting scenario could not be detected. This is linked to the fact that we are very close to a linear model scenario, meaning that in the compressed space the model still has enough flexibility to fit the data. This motivated a deeper testing of the framework, extending it to extra degrees of freedom as follows.

\begin{figure*}
    \centering
    \includegraphics[width=0.325\textwidth, clip=true, trim = 0.4cm 0.5cm 0.35cm 0.3cm]{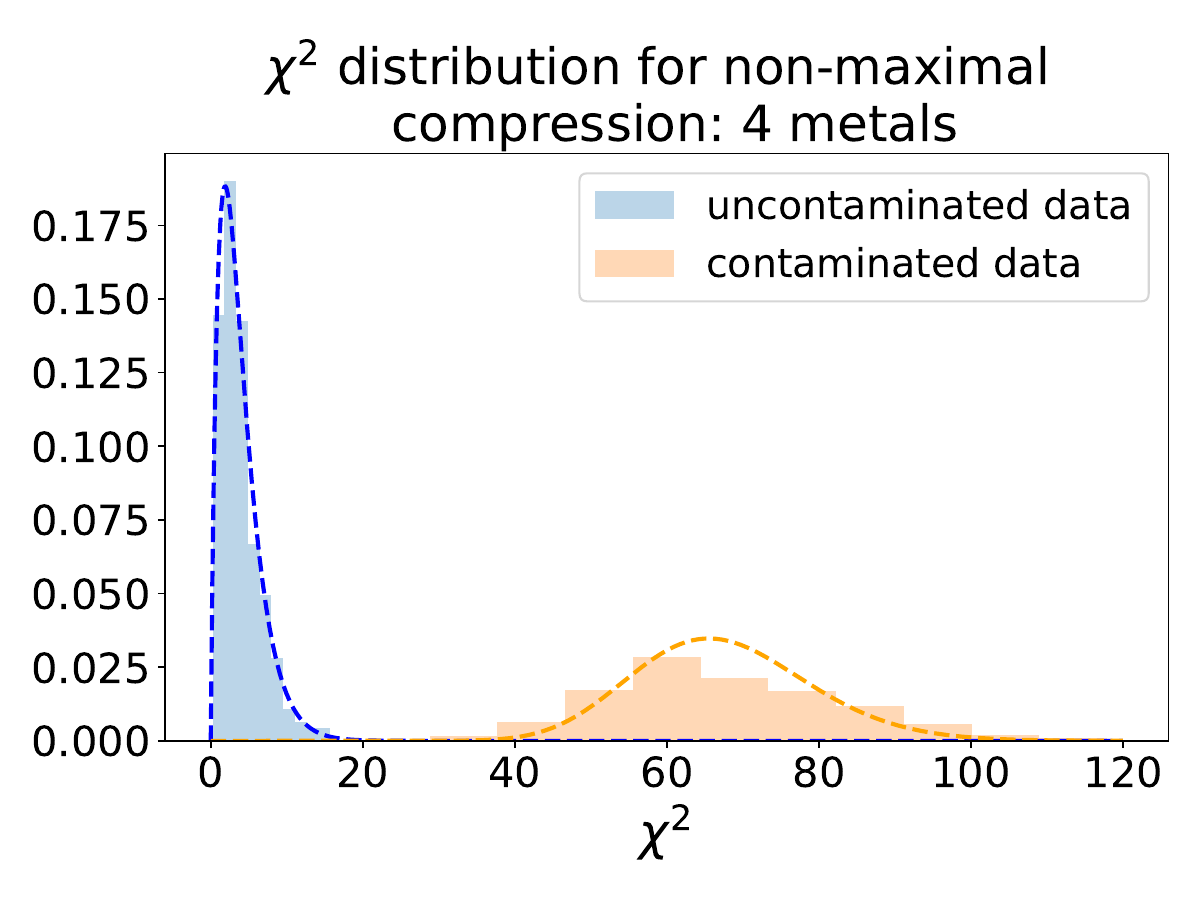}
    \includegraphics[width=0.325\textwidth, clip=true, trim = 0.4cm 0.5cm 0.35cm 0.3cm]{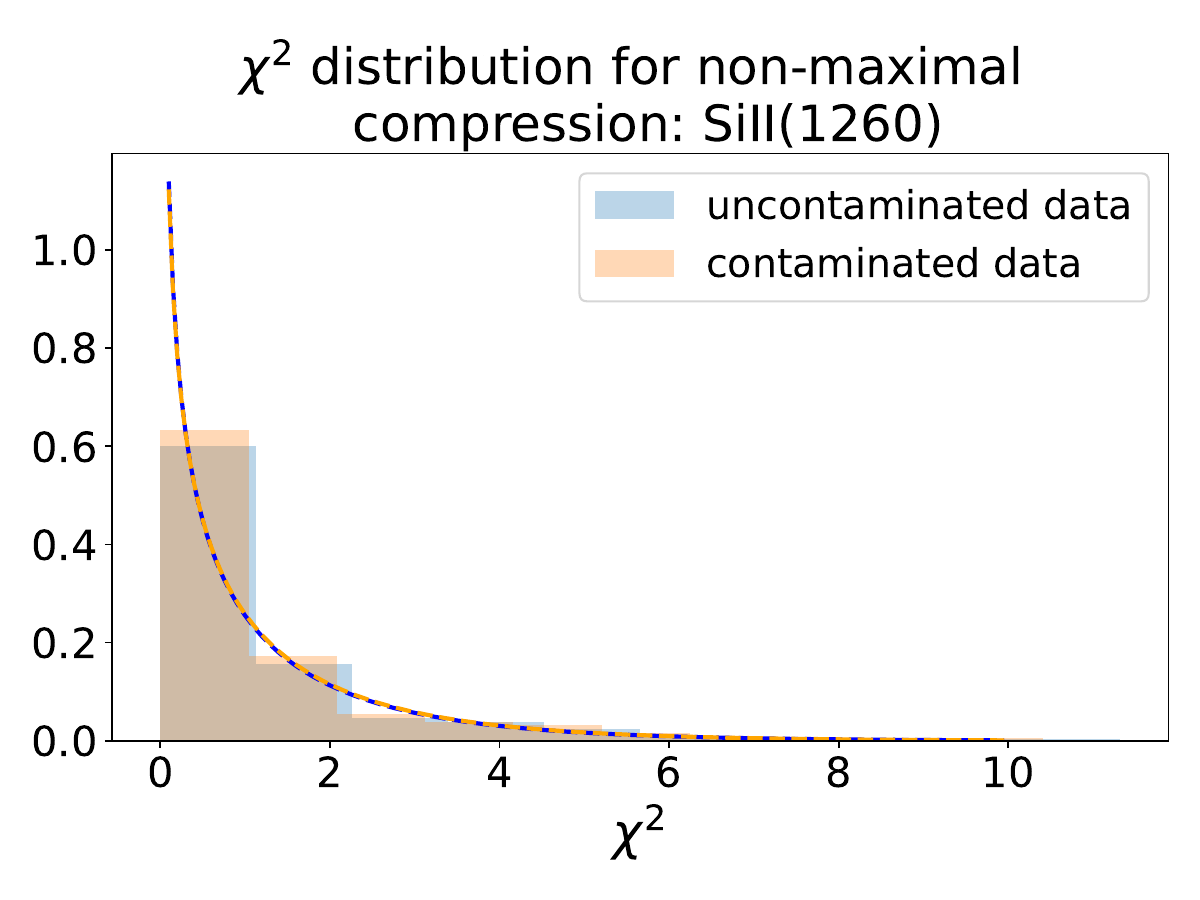}
    \includegraphics[width=0.325\textwidth, clip=true, trim = 0.25cm 0.5cm 0.35cm 0.3cm]{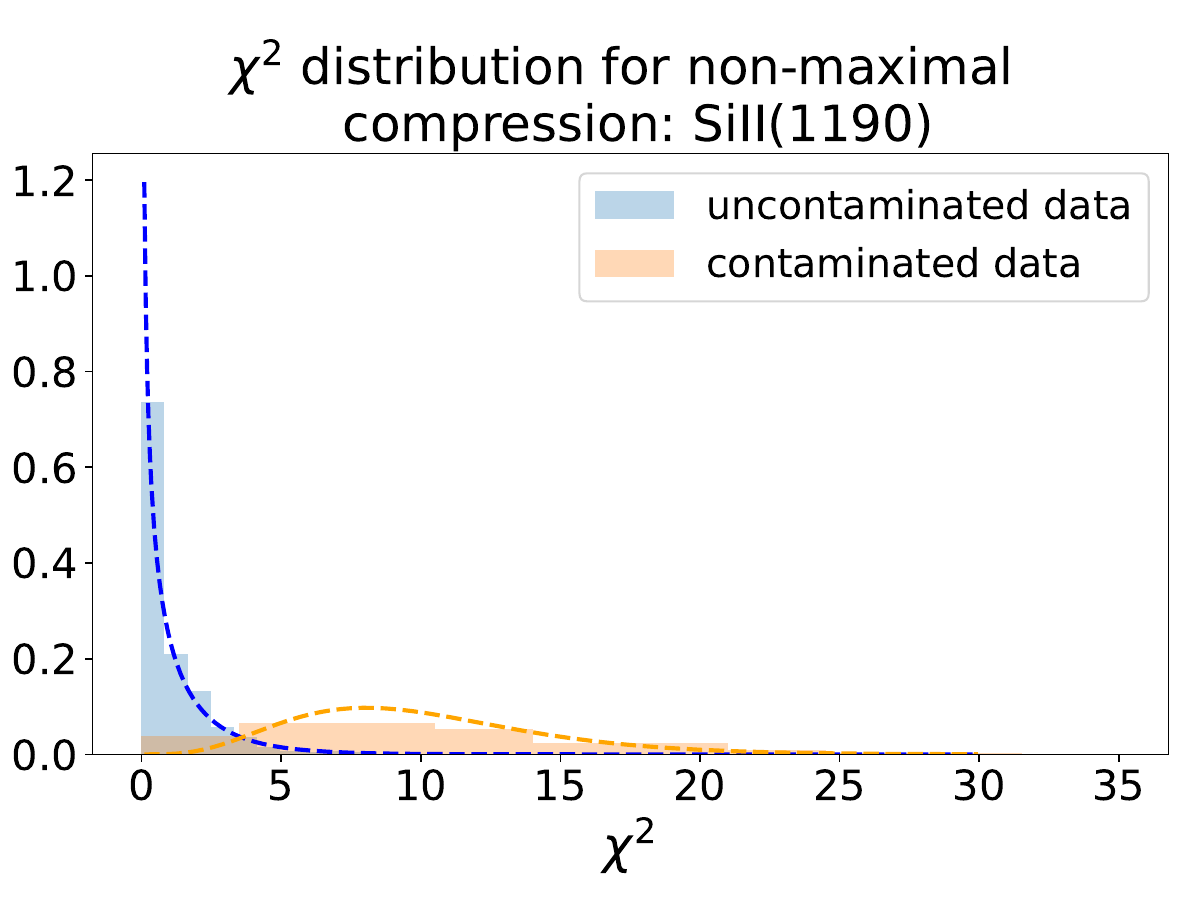}
    \caption{Normalized $\chi^2$ histograms for the three \textit{non-maximal compression} cases presented in Sect. \ref{subsect:nonmaximal_compression}: starting from the left, all four metals, SiII(1260) and SiII(1190) were used to build extra degrees of freedom. In blue the histograms and $\chi^2$ distributions for the uncontaminated data, orange for contaminated. The corresponding $\chi^2$ distributions (dashed lines) are generated assuming as number of degrees of freedom the mean of the histogram distributions.
    The first set of histograms, that relates to all four extra degrees of freedom, present a strong shift between the orange and the blue distributions: their corresponding means are 3.89 and 67.51. In the SiII(1260) case, both distributions have a mean of $\sim 1.1$, while in the SiII(1190), the mean for the contaminated case is $1.01$, against $10.04$ in the contaminated case.}
    \label{fig:case_cd_metals}
\end{figure*}

\subsection{Non-maximal compression}\label{subsect:nonmaximal_compression}

Given the problem highlighted in the \textit{maximal} framework, we tested the pipeline in a \textit{non-maximal compression} case, where the extra degrees of freedom are given by the metals contaminating the data. Namely, the \textit{maximal} summary statistics is now extended to include $\boldsymbol{t}_{\mathrm{extra}}=\{ t_{b_{\eta, \mathrm{SiII(1260)}}}, t_{b_{\eta, \mathrm{SiII(1193)}}}, t_{b_{\eta, \mathrm{SiIII(1207)}}}, t_{b_{\eta, \mathrm{SiII(1190)}}} \}$. Still, metals will not be included in the likelihood modelling. This means that if the quantities of reference here are the compressed data vector
\begin{equation}
    \boldsymbol{t} = \nabla \boldsymbol{\mu}_*^T \boldsymbol{C}^{-1} (\boldsymbol{d}-\boldsymbol{\mu}_*) \; ,
\end{equation}
the compressed model
\begin{equation}
    \boldsymbol{\mu_t} = \nabla \boldsymbol{\mu}_*^T \boldsymbol{C}^{-1} (\boldsymbol{\mu(\theta)}-\boldsymbol{\mu}_*) \; ,
\end{equation}
and they enter the $\chi^2$ as per
\begin{equation}
    \chi^2(\boldsymbol{\theta}) = [\boldsymbol{t}-\boldsymbol{\mu}_{\boldsymbol{t}}(\boldsymbol{\theta})]^T \boldsymbol{F}^{-1} [\boldsymbol{t}-\boldsymbol{\mu}_{\boldsymbol{t}}(\boldsymbol{\theta})] \; ,
\end{equation}
the fiducial model $\boldsymbol{\mu}_*$ and its gradient will now include contaminants, whereas $\boldsymbol{\mu(\theta)}$ will not and $\boldsymbol{d}$ will either be contaminanted or uncontaminated data depending on the population used to build the $\chi^2$ statistics. Now $\boldsymbol{t} = \{ \boldsymbol{t}_{\mathrm{max}},\boldsymbol{t}_{\mathrm{extra}} \}$.  The length of the compressed data vector is ten, where the first six components refer to the sampled parameters, with a remainder of four components, which are fixed and constitute our extra degrees of freedom.  Under the approximation that the mean of a $\chi^2$ distribution indicates the number of degrees of freedom of the problem, we would expect that mean to be at least equal to the number of extra degrees of freedom we added. In our case, we expect that for the uncontaminated case, for which we know the modelling is good, the mean will be close to 4 (four metals). We want to test whether in this case a bad fit to the contaminated data is apparent as a mean $\chi^2$ significantly larger than 4.


The $\chi^2$ histograms are shown in the left panel of Fig.~\ref{fig:case_cd_metals}: the mean values for the uncontaminated and contaminated cases are respectively 3.89 and 67.51. Considering a $\chi^2$ with number of degrees of freedom equal to 4, the p-values for the two means are respectively 0.4 and $10^{-14}$: the bad fit in the contaminated case has emerged.

We further experimented over the addition of metals and we considered adding a single extra degree of freedom at a time, associated to either one of the following metals: the SiII(1260) and the SiII(1190). The resulting $\chi^2$ histograms are shown in the middle and right panels of Fig.~\ref{fig:case_cd_metals}, respectively. These two metal lines were chosen because of how differently they affect the data: while the SiII(1260) contamination happens around the BAO scale along the line-of-sight, the SiII(1190) contributes to the peak at $\sim 60 \mathrm{Mpc}/h$. 
We run the same exact experiment and find that the addition of $t_{b_{\eta, \mathrm{SiII(1190)}}}$ does bring out the bad fit, while the other does not. Specifically, the two $\chi^2$ distributions when the extra degree of freedom is given by $b_{\eta, \mathrm{SiII(1260)}}$ have a mean of $\sim 1$, again equal to the number of degrees of freedom, but they cannot be distinguished. The p-values for both distributions, assuming one degree of freedom, are all above a threshold of 0.01. Both distributions are indicative of an acceptable fit. On the contrary, adding the extra compressed component related to SiII(1190) results in having a mean $\chi^2$ of 1.01 in the uncontaminated case and 10.04 in the contaminated one, with corresponding p-values of 0.3 and $10^{-3}$ if we consider a target $\chi^2$ distribution of one degree of freedom. This perhaps is indicative about the fact that in order to capture a bad fit, adding extra degrees of freedom is not enough: these extra degrees of freedom must be informative about features not captured by the core set of parameters. The SiII(1260) affects the model at scales of the correlation function which are on top of the BAO peak, which we model for, whereas SiII(1190) effectively adds information on a feature which is completely unmodelled.

In light of this, a possible solution is to add some extra degrees of freedom to the \textit{maximal} compression vector, which are designed to be orthogonal to the already known components in the compressed space. This would allow the extra flexibility, that is not captured in the model, to highlight for a bad fit in the compressed framework. This is an interesting problem which is left for future work. However, a similar solution has already been implemented in the context of MOPED \citep{Heavens:2020spq}, specifically to allow new physics to be discovered.

Not modelling the SiII(1260) line in the uncompressed traditional framework does not result in any bad fit, which makes this an example of systematics hidden in the large original data vector. At the same time, the fact that the SiII(1260) test in the compressed framework fails to show a bad fit at the level of the $\chi^2$ is quite problematic, given this metal line is one of the primary contaminants we have to be careful of in BAO measurement, affecting the peak's scale. The worry is then that, despite constructing an extended framework, there is a chance that some systematics hiding in the signal could be missed. This effectively means that in order to apply data compression, the underlying physics must be already well known to a good extent. 
Because some systematics could be either hard to model or to detect, in this example, we deliberately assumed we had no knowledge about known systematics, where in principle we could have also marginalized over them \citep{nuisance_hardened}.

\section{Robustness to parameter non-linearities}\label{sect:fiducials_ensemble}

Each component of the score-compressed data vector relates to a specific model parameter, as per Eq.~(\ref{eqn:score_formula}), via the gradient. Throughout the analysis, the BAO parameters proved to be a source of non-linearities in relation to their summary statistics components (see Fig.~\ref{fig:score_apcompressed}), sometimes resulting in a multi-peaked posterior distribution. With the intent of mitigating this effect, we were forced to impose a tight prior on both $\{ \alpha_{\parallel},\alpha_{\perp} \}$, which reduces the generalizability of the approach.

\begin{figure}
    \centering
    \includegraphics[width=0.33\textwidth, clip=true, trim = 0.5cm 0.5cm 0.35cm 0.3cm]{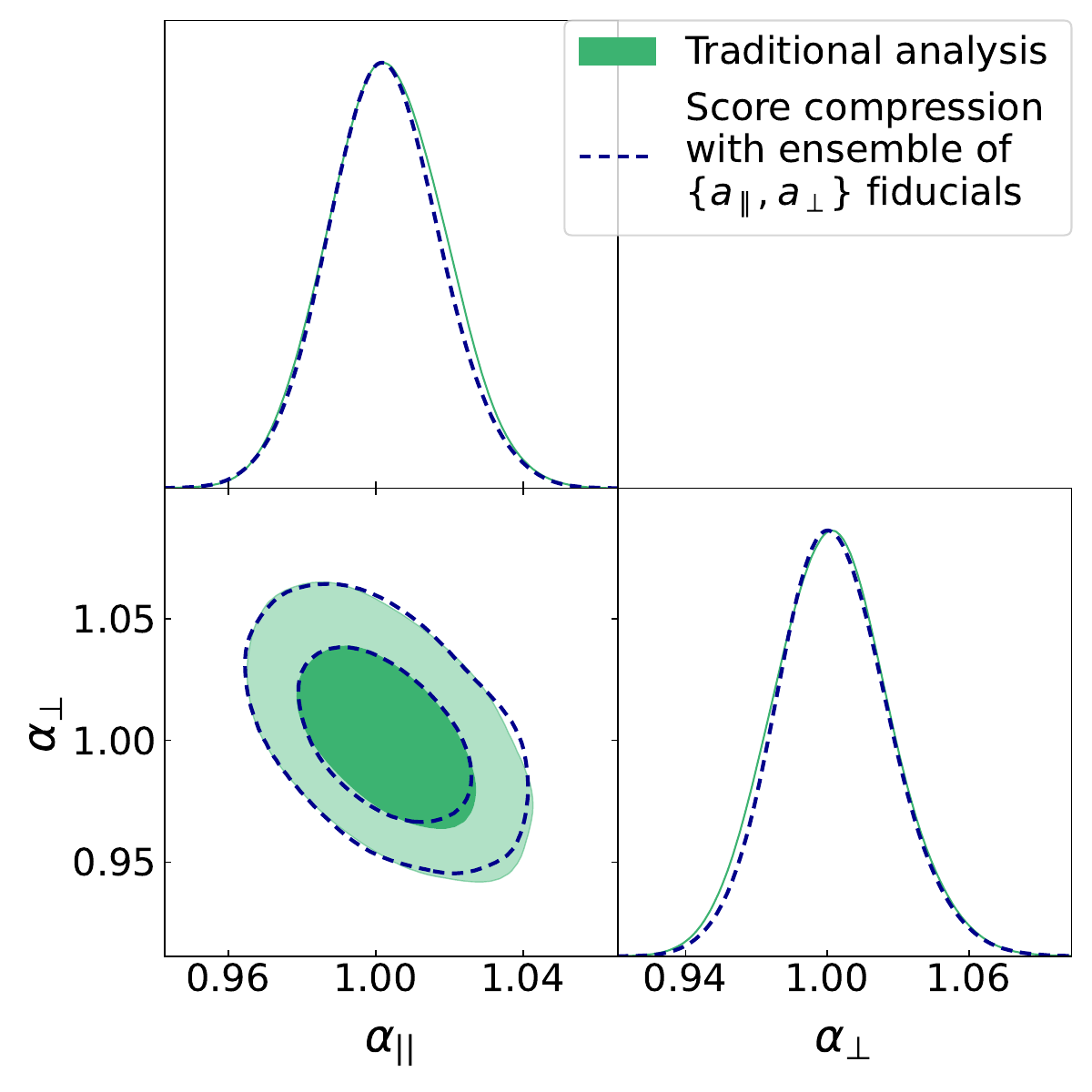}
    \caption{Triangle plots of the BAO parameters of interest $\{\alpha_{\parallel}, \alpha_{\perp}\}$ for one set of the Ly$\alpha$ auto- and cross- mock correlations, with relaxed priors. The green filled contours refer to the results obtained performing the inference using the full uncompressed data vector, which we denote as ‘Traditional analysis’, while the blue dashed refer to the compressed analysis results, denoted as ‘Score compression analysis’. The framework of the latter is extended here to the assumption of multiple fiducial values for $\{\alpha_{\parallel}, \alpha_{\perp}\}$ when performing the compression, namely $[ \{ \alpha_{\parallel} = 1.00 ,\alpha_{\perp} = 1.01 \}, \{ \alpha_{\parallel} = 0.8 ,\alpha_{\perp} = 1.2 \}, \{ \alpha_{\parallel} = 1.2 ,\alpha_{\perp} = 0.8 \}, \{ \alpha_{\parallel} = 1.3 ,\alpha_{\perp} = 0.7 \}, \{ \alpha_{\parallel} = 0.9 ,\alpha_{\perp} = 1.1 \} ]$.}
    \label{fig:score_fiducialensemble}
\end{figure}

Based on the work of \cite{Protopapas}, we explore extensions to the algorithm by considering an ensemble of fiducial values of the BAO parameters to compute the score-compressed vector components related to $\{ \alpha_{\parallel},\alpha_{\perp} \}$. For any extra set of BAO parameters $\{ \alpha^{\mathrm{extra}}_{\parallel},\alpha^{\mathrm{extra}}_{\perp} \}$, we introduce two extra summary statistics components:
\begin{align}
    \boldsymbol{t}^{\mathrm{extra}}_{\alpha_{\parallel}} = \nabla_{\alpha_{\parallel}} \boldsymbol{\mu}_{\mathrm{extra}}^T \boldsymbol{C}^{-1} (\boldsymbol{d}-\boldsymbol{\mu}_{\mathrm{extra}}) \; , \\
    \boldsymbol{t}^{\mathrm{extra}}_{\alpha_{\perp}} = \nabla_{\alpha_{\perp}} \boldsymbol{\mu}_{\mathrm{extra}}^T \boldsymbol{C}^{-1} (\boldsymbol{d}-\boldsymbol{\mu}_{\mathrm{extra}}) \; ,
\end{align}
where $\boldsymbol{\mu}_{\mathrm{extra}}$ is the model evaluated at $\{ \boldsymbol{\alpha}^{\mathrm{extra}}_{\parallel},\boldsymbol{\alpha}^{\mathrm{extra}}_{\perp} \}$, keeping the previously defined fiducial values for the other parameters. As these extra components effectively represent an extension of the compressed dataset, the Fisher matrix in Eq.~(\ref{eqn:fisher}) will also be expanded to include $[\nabla_{\alpha_{\parallel,\perp}} \boldsymbol{\mu}_{\mathrm{extra}}]^T \boldsymbol{C}^{-1}[\nabla_{\alpha_{\parallel,\perp}}^{T} \boldsymbol{\mu}_{\mathrm{extra}}]$. We test this extension on the same mock that was used to test the subsampling covariance matrix in Sect.~\ref{sect:results_testcovariance}, and results are presented in Fig.~\ref{fig:score_fiducialensemble}, imposing a physically motivated uniform prior $[0.65, 1.35]$ for both $\alpha_{\parallel}$ and $\alpha_{\perp}$. The ensemble of extra fiducials is given by the set $[ \{ \alpha_{\parallel} = 0.8 ,\alpha_{\perp} = 1.2 \}, \{ \alpha_{\parallel} = 1.2 ,\alpha_{\perp} = 0.8 \}, \{ \alpha_{\parallel} = 1.3 ,\alpha_{\perp} = 0.7 \}, \{ \alpha_{\parallel} = 0.9 ,\alpha_{\perp} = 1.1 \} ]$, in addition to the original $\{ \alpha_{\parallel} = 1.00 ,\alpha_{\perp} = 1.01 \}$ (see Tab.~\ref{tab:score_fulltable}). From Fig.~\ref{fig:score_fiducialensemble} it can be seen that the constraining power on the BAO parameters between the traditional and compressed methods match. This same result is also true for the other parameters, not shown here.

We tested the extension in terms of generalizability by progressively adding extra points to the ensemble, with reasonable spread, and found that with an ensemble of three to four extra fiducial sets of BAO parameters the algorithm is able to effectively get rid of the secondary posterior peaks and increase the accuracy of the measurement. Hence, the assumption of multiple fiducials for the BAO parameters, for which we had to impose a tight prior, enables us to relax the prior constraints.

\section{Application to real data}\label{sect:score_eboss}

\begin{figure*}
    \centering
    {\includegraphics[width=0.64\textwidth, clip=true, trim = 0.47cm 0.5cm 0.35cm 0.3cm]{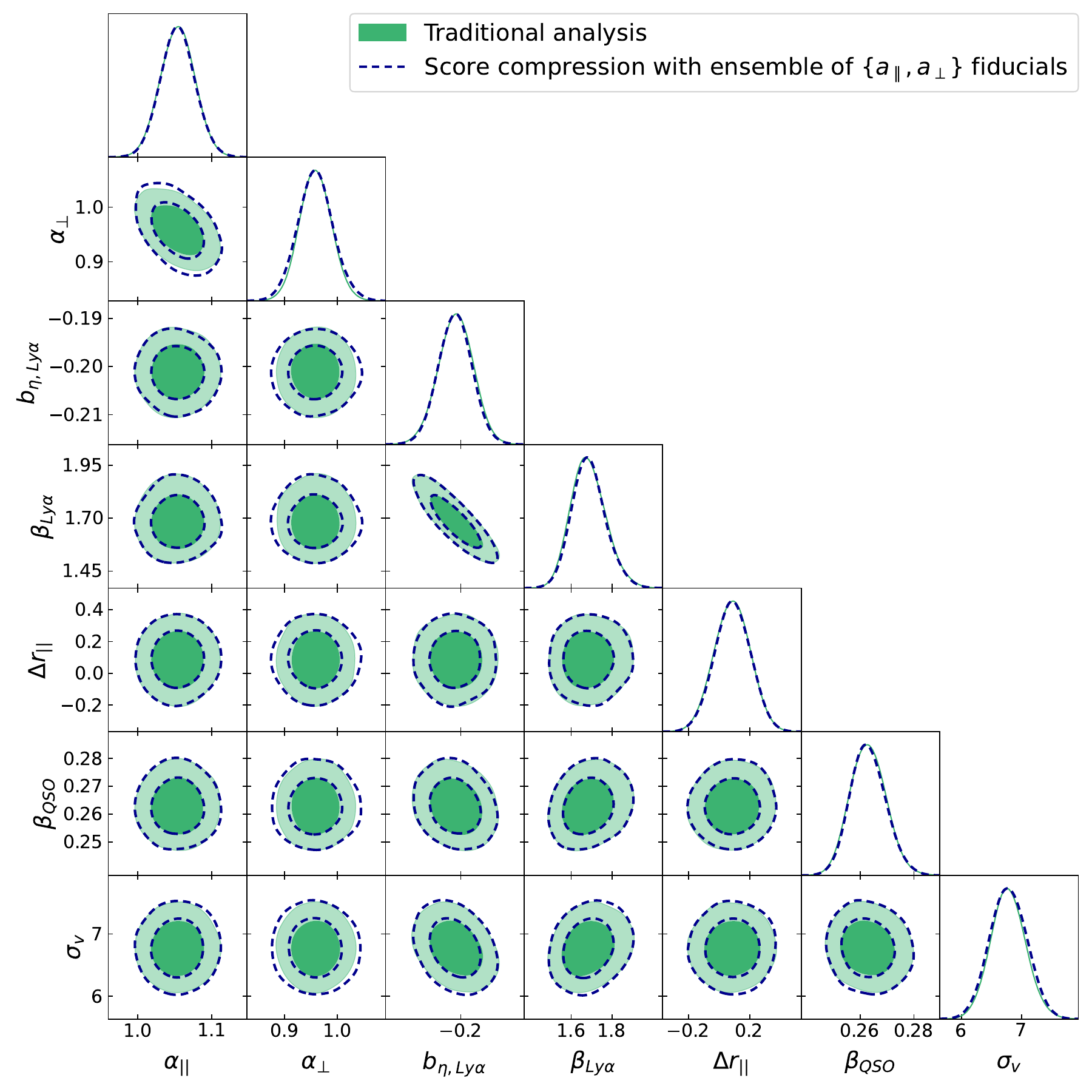}}
    \caption{Triangle plot for fits to the real eBOSS DR16 data Ly$\alpha$ auto- and cross- correlations, using the traditional approach (filled green) and the score compression framework (dashed blue) extended to include extra fiducial values of the BAO parameters at $[ \{ \alpha_{\parallel} = 0.8 ,\alpha_{\perp} = 1.2 \}, \{ \alpha_{\parallel} = 1.2 ,\alpha_{\perp} = 0.8 \}, \{ \alpha_{\parallel} = 1.3 ,\alpha_{\perp} = 0.7 \}, \{ \alpha_{\parallel} = 0.9 ,\alpha_{\perp} = 1.1 \} ]$. The results shown here are for the standard parameters $\{\alpha_{\parallel}, \alpha_{\perp},  b_{\eta, \rm{Ly}\alpha}, \beta_{\rm{Ly}\alpha}, $ $\Delta r_{\parallel}, \beta_{\rm{QSO}}, \sigma_{\rm{v}} \}$.}
    \label{fig:eboss_posteriors}
\end{figure*}

The score compression framework has so far been tested on realistic mocks, hence it is straightforward to apply this same algorithm to real eBOSS DR16 Ly$\alpha$ data, for which we refer to \cite{Bourboux_Rich_Font-Ribera_Agathe_Farr_Etourneau_Goff_Cuceu_Balland_Bautista_et_al._2020}. The set of nuisance parameters is now extended to also include the contamination from carbon absorbers, the systematic quasar redshift error $\Delta r_{\parallel}$, the quasar radiation strength $\xi_0^{\mathrm{TP}}$ and the sky-subtraction parameters $A_{\mathrm{sky,Ly}\alpha}$ and $\sigma_{\mathrm{sky,Ly}\alpha}$. The results presented in Sect.~\ref{sect:fiducials_ensemble} motivate a direct test of the whole extended framework, which gets rid of the tight prior, to the real data. The ensemble of BAO parameter fiducial values is given by the set of $\{ \alpha_{\parallel} = 1.05,\alpha_{\perp} = 0.96 \}$ --- which are the best fit values obtained through the traditional analysis --- and $[ \{ \alpha_{\parallel} = 0.8 ,\alpha_{\perp} = 1.2 \}, \{ \alpha_{\parallel} = 1.2 ,\alpha_{\perp} = 0.8 \}, \{ \alpha_{\parallel} = 1.3 ,\alpha_{\perp} = 0.7 \}, \{ \alpha_{\parallel} = 0.9 ,\alpha_{\perp} = 1.1 \} ]$, which were found to be effective in Sect.~\ref{sect:fiducials_ensemble}. The fiducial values of the other parameters are set to the best fit found with the standard uncompressed analysis. In Fig.~\ref{fig:eboss_posteriors}, we present the agreement of the extended framework against the traditional approach at the level of $\{\alpha_{\parallel}, \alpha_{\perp},  b_{\eta, \rm{Ly}\alpha}, \beta_{\rm{Ly}\alpha}, $ $\Delta r_{\parallel}, \beta_{\rm{QSO}}, \sigma_{\rm{v}} \}$. The nuisance parameters are also found to be in excellent agreement.

\section{CONCLUSIONS}\label{sect:score_conclusions}

Standard analyses of the Lyman-$\alpha$ (Ly$\alpha$) forest correlation functions focus on a well localized region, which corresponds to the baryon acoustic oscillations (BAO) peak. However, these correlation functions usually have dimensions of 2500 or 5000, which means the cosmological signal is extracted from a small subset of bins.  This means that reducing the dimensionality of the data vector, while retaining the information we care about, could be a step forward in optimizing the analysis. At the same time, as extensively explained in Sect.~\ref{sect:score_method}, the covariance matrix $\mathbf{C}$ used for Ly$\alpha$ correlations analyses is estimated via sub-sampling. However, the dimensionality of the correlation functions is much larger than the number of data samples used to estimate the covariance. Reducing the dimensionality of the data vector to $\mathcal{O}(10)$ allows for a reliable estimate of the covariance matrix. Given these premises, the goal of this work is to apply and explore a data compression algorithm for realistic Ly$\alpha$ auto- and cross-correlation functions.

We reduced the dimensionality of the data vector to a set of summary statistics $\mathbf{t}$ using score compression. We assume a Gaussian likelihood, test for its validity, and show that this assumption is preserved in the compressed space as well, as the compression is a linear transformation. In the compressed space the covariance can be either given by the mapped traditional covariance or by a covariance estimated directly in such a space.

We tested the compressed framework against the traditional approach at the posterior level, when using the original covariance $\mathbf{C}$, and found the two of them agree, and no bias is introduced. We then showcased a test example of covariance matrix evaluation in the compressed space, which is a key benefit of the approach, enabling a comparison to the covariance matrix obtained in the traditional sub-optimal framework. Because of non-linear relationship between the BAO parameters and their summary statistics components, throughout the analysis we adopted a tight prior on $\{ \alpha_{\parallel},\alpha_{\perp} \}$. Later in the analysis, with the aim of increasing the generalizability of the approach, while relaxing the prior constraint, we successfully tested extensions to the framework by assuming an ensemble of fiducial values for these problematic parameters.

We then further examined the compressed framework, by testing the inference against unmodelled effects and we find that if any information about the unmodelled features in the correlation function is not captured by the compressed data vector $\boldsymbol{t}$, this can potentially lead to biases, which do not emerge at the level of the $\chi^2$ goodness of fit test. Hence, we advise against performing goodness of fit tests in compressed space, unless the compressed vector is extended to include extra degrees of freedom, analogous to what is done in \cite{Heavens:2020spq}. Extending the framework in this sense is left for future work.

We applied our extended compression framework to DR16 data from the Extended Baryon Oscillation Spectroscopic Survey and demonstrated that the posterior constraints are  accurately recovered without loss of information. 
A step change in constraining power, and thus accuracy requirements, is expected for forthcoming Ly$\alpha$ cosmology analyses by the on-going DESI experiment \citep[see e.g.,][]{gordon20233d}, which will observe up to 1 million high-redshift quasars with $z > 2$.
Optimal data compression as proposed in this work will facilitate these analyses through inference that is complementary to the traditional approach and through additional consistency and validation tests.

\section*{Acknowledgements}

We thank Alan Heavens, Niall Jeffrey and Peter Taylor for helpful discussions. This work was partially enabled by funding from the UCL Cosmoparticle Initiative. AC acknowledges support provided by NASA through the NASA Hubble Fellowship grant HST-HF2-51526.001-A awarded by the Space Telescope Science Institute, which is operated by the Association of Universities for Research in Astronomy, Incorporated, under NASA contract NAS5-26555. BJ acknowledges support by STFC Consolidated Grant ST/V000780/1. SN acknowledges support from an STFC Ernest Rutherford Fellowship, grant reference ST/T005009/2. AFR acknowledges support by the programme Ramon y Cajal (RYC-2018-025210) of the Spanish Ministry of Science and Innovation and from the European Union’s Horizon Europe research and innovation programme (COSMO-LYA, grant agreement 101044612). IFAE is partially funded by the CERCA program of the Generalitat de Catalunya.
For the purpose of open access, the authors have applied a creative commons attribution (CC BY) licence to any author-accepted manuscript version arising.

\section*{Data Availability}

The code is publicly available at the `compression' branch of \url{https://github.com/andreicuceu/vega.git}. The data underlying this article will be shared on reasonable request to the corresponding author.
 



\bibliographystyle{mnras}
\bibliography{main} 




\appendix
\section{Full results for the mock-to-mock covariance test}\label{appendix}

\begin{figure*}
    \centering
    \begin{tikzpicture}
      \node (img1) {\includegraphics[width=0.78\textwidth, clip=true, trim = 0.4cm 0.5cm 0.35cm 0.3cm]{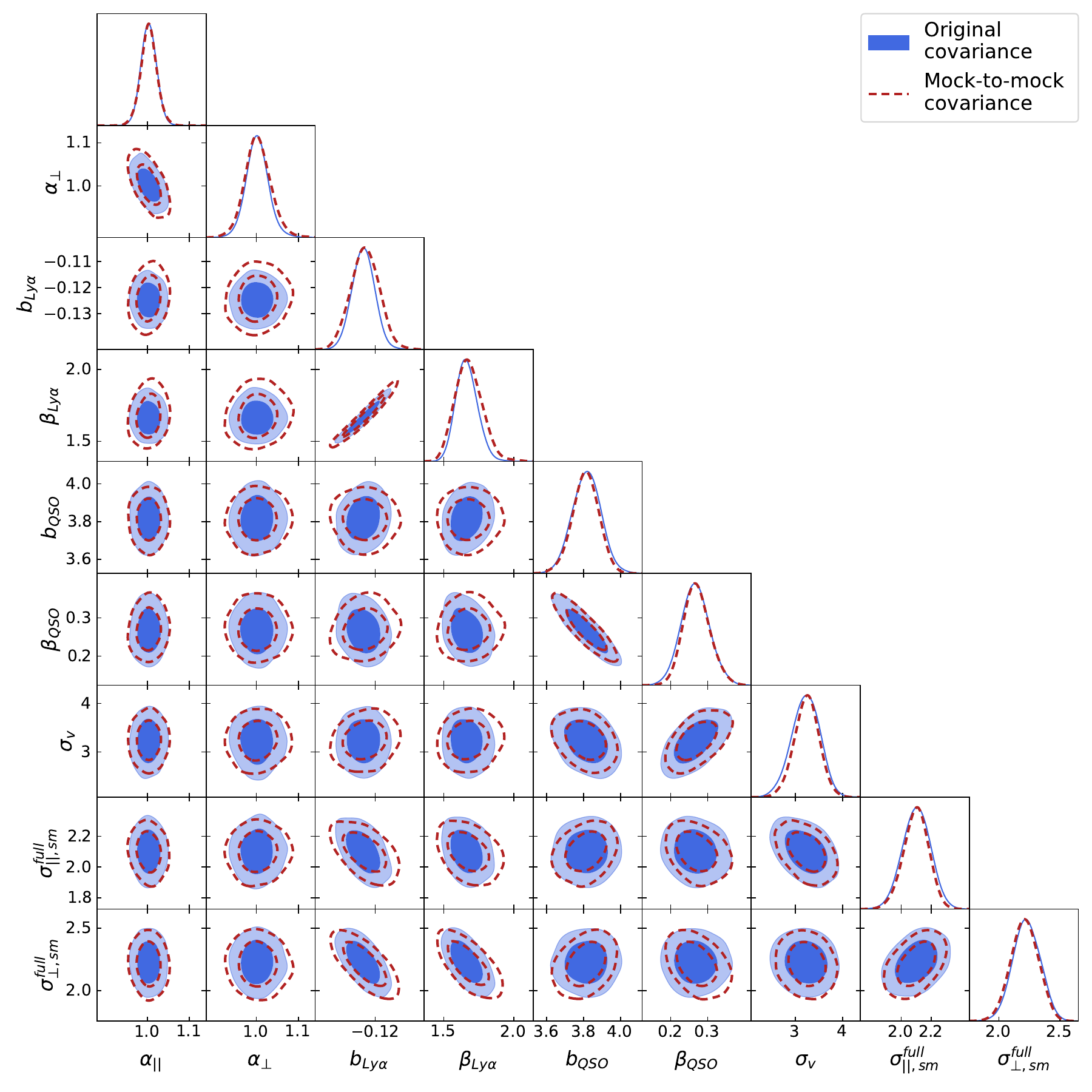}};
      \node (img2) at (6.2, 4.6) {\includegraphics[width=0.55\textwidth, clip=true, trim = 0.4cm 0.5cm 0.35cm 0.3cm]{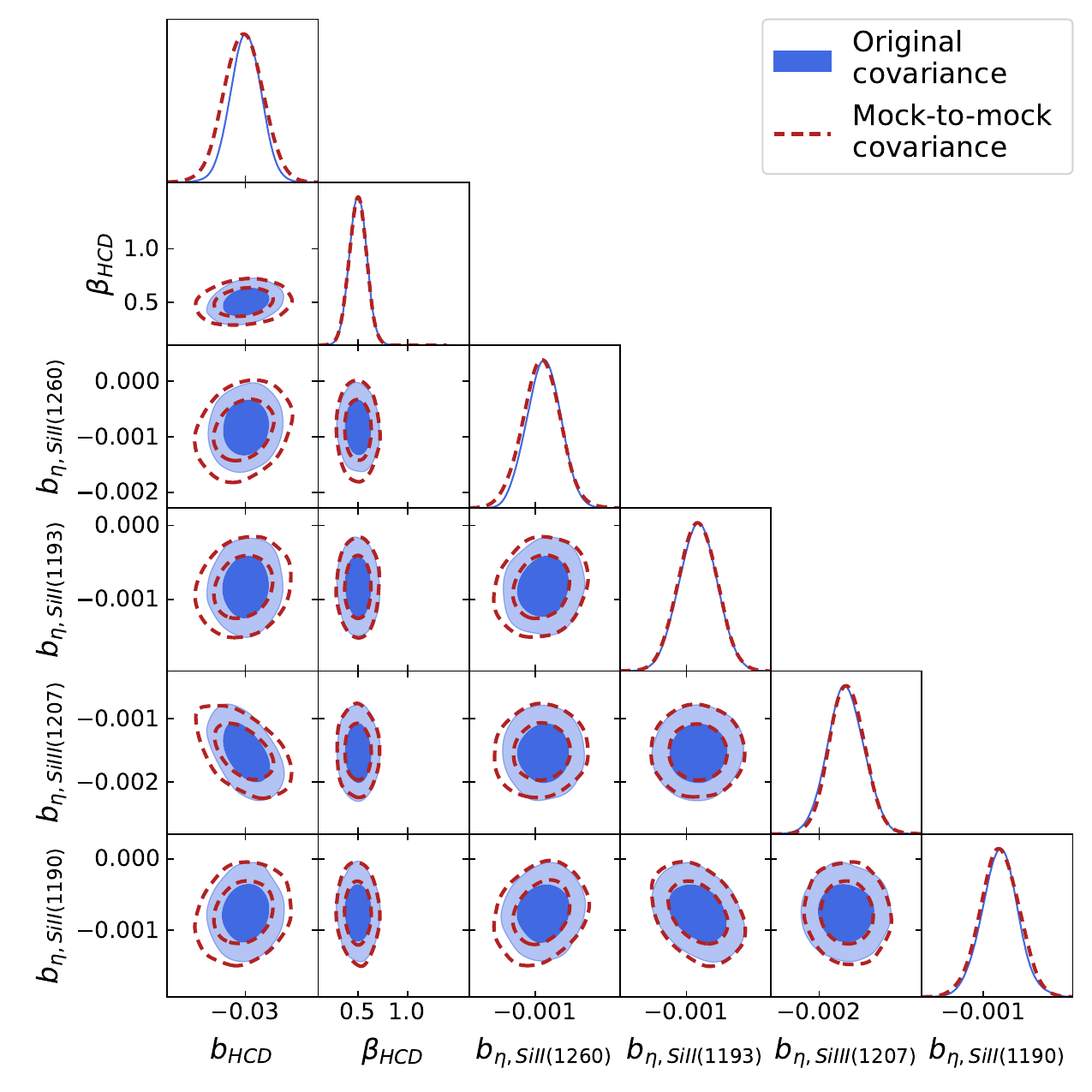}};
    \end{tikzpicture}
    \caption{Triangle plots of the parameters of interest for one set of the Ly$\alpha$ auto- and cross- correlation mocks. Results are split, for presentation purposes only, into the set of standard parameters $\{\alpha_{\parallel}, \alpha_{\perp},  b_{\rm{Ly}\alpha}, \beta_{\rm{Ly}\alpha}, b_{\rm{QSO}}, \beta_{\rm{QSO}}, \sigma_{\rm{v}}, \sigma_{\parallel}, \sigma_{\perp} \}$ (lower left panel) and contaminants parameters $\{ b_{\eta, \mathrm{SiII(1260)}}, b_{\eta, \mathrm{SiII(1193)}}, b_{\eta, \mathrm{SiIII(1207)}}, b_{\eta, \mathrm{SiII(1190)}}, b_{\mathrm{HCD}}, \beta_{\mathrm{HCD}} \}$ (upper right panel). The blue filled contours refer to the results obtained performing the inference using the original covariance matrix $\boldsymbol{C}$ mapped into the compressed space (the Fisher matrix) in the likelihood function, and hence are denoted as ‘Original covariance’. On the other hand, the red dashed results, denoted as 'Mock-to-mock covariance', refer to the case in which the mock-to-mock covariance matrix is used instead, while adopting a t-distribution likelihood. }
    \label{fig:score_cov_posterior}
\end{figure*}

We here present in Fig.~\ref{fig:score_cov_posterior} the full set of results from the mock-to-mock covariance test, presented in Sect.~\ref{sect:results_testcovariance}, against the contours obtained using the original covariance in the compressed framework. Numerical values are reported in Tab.~\ref{tab:score_fulltable}. The contours agree well with each other. The most striking enlargements of the posteriors are visible for the parameters $\{\alpha_{\perp},  b_{\rm{Ly}\alpha}, \beta_{\rm{Ly}\alpha}, b_{\mathrm{HCD}}\}$. Because the `Original covariance' setup has been shown to agree with the standard analysis in Sect.~\ref{sect:score_results}, this comparison automatically becomes a comparison to the standard approach. 


\bsp	
\label{lastpage}
\end{document}